\documentclass[11pt]{article}

\usepackage{jheppub}
\usepackage{hyperref}
\usepackage{footmisc}
\usepackage{booktabs}
\usepackage{comment}
\usepackage{tabularx}

\hypersetup{
  colorlinks,
  hypertexnames=false,
  urlcolor=blue,
  linkcolor=magenta,
  citecolor=cyan
}

\usepackage{bm, bbm, bbold}
\usepackage{latexsym}
\usepackage{dcolumn}
\usepackage{amsmath,amsfonts,amssymb, mathrsfs, amsthm}
\usepackage[T1]{fontenc}

\usepackage{graphicx,epsfig}
\usepackage{environ}
\usepackage{mathtools}
\usepackage{braket}
\usepackage{fancyhdr}
\usepackage{hyperref}
\usepackage{graphicx,epstopdf}
\usepackage{tikz}
\usepackage{float}
\usepackage{framed}
\usepackage{soul}
\usetikzlibrary{positioning,decorations.markings, decorations.pathmorphing, calc}
\tikzset{snake it/.style={decorate, decoration=snake}}
\usetikzlibrary{angles,quotes}

\usepackage{stmaryrd}
\usepackage{slashed}
\usepackage{array,multirow}

\usepackage[framemethod=default]{mdframed}
\newmdenv[skipabove=7pt,
skipbelow=7pt,
rightline=false,
leftline=false,
topline=false,
bottomline=false,
backgroundcolor=gray!10,
linecolor=gray,
innerleftmargin=5pt,
innerrightmargin=5pt,
innertopmargin=5pt,
innerbottommargin=5pt,
leftmargin=0cm,
rightmargin=0cm,
linewidth=4pt]{eBox}

\newcommand{\be}{\begin{equation}}
\newcommand{\ee}{\end{equation}}
\newcommand{\bes}{\begin{equation*}}
\newcommand{\ees}{\end{equation*}}

\NewEnviron{derivation}{
\begin{framed}
\begin{center}
{\bf-: Derivation :-}\\
\end{center}
  \BODY

\end{framed}
}

\NewEnviron{eqn}{
\begin{align}
\begin{split}
  \BODY
\end{split}
\end{align}
}

\NewEnviron{eqn*}{
\begin{align*}
\begin{split}
  \BODY
\end{split}
\end{align*}
}

\newcommand{\red}[1]{{\color{red}#1}}

\definecolor{darkindigo}{RGB}{75,70,140}

\usepackage{cancel}
\usepackage{tikz, pgf}
\usetikzlibrary{shapes.misc}
\usetikzlibrary{shapes}
\usetikzlibrary{fit}
\usetikzlibrary{arrows}

\usetikzlibrary{decorations.pathmorphing}	% For Feynman Diagrams
\usetikzlibrary{decorations.markings}
\tikzset{
    sugra/.style={decorate, decoration={snake}, draw=black},
    scalarphi/.style={dashed,draw=black, postaction={decorate},
        },
    hwbou/.style={draw=blue, postaction={decorate}, ultra thick
        },
    vector/.style={draw=blue,decorate, decoration={snake}, draw},
	provector/.style={decorate, decoration={snake,amplitude=2.5pt}, draw},
	antivector/.style={decorate, decoration={snake,amplitude=-2.5pt}, draw},
   	 fermion/.style={draw=cyan, postaction={decorate},
        decoration={markings,mark=at position .55 with {\arrow[draw=black]{>}}}},
    fermionbar/.style={draw=cyan, postaction={decorate},
        decoration={markings,mark=at position .55 with {\arrow[draw=black]{<}}}},
    fermionnoarrow/.style={draw=black},
    gluon/.style={decorate, draw=red,
        decoration={coil,amplitude=4pt, segment length=5pt}},
    scalar/.style={dashed,draw=black, postaction={decorate},
        decoration={markings,mark=at position .55 with {\arrow[draw=black]{>}}}},
    scalarbar/.style={dashed,draw=black, postaction={decorate},
        decoration={markings,mark=at position .55 with {\arrow[draw=black]{<}}}},
    electron/.style={draw=black, postaction={decorate},
        decoration={markings,mark=at position .55 with {\arrow[draw=black]{>}}}},
    scalarnoarrow/.style={dashed, draw=black},
    electron/.style={draw=black, postaction={decorate},
        decoration={markings, mark=at position .55 with {\arrow[draw=black]{>}}}},
	bigvector/.style={decorate, decoration={snake, amplitude=4pt}, draw},
    photon/.style={draw=red, decorate, decoration={zigzag}, draw},
    higgs/.style={dashed, draw=black, postaction={decorate},
        },	
        goldstone/.style={draw=brown, postaction={decorate},
        },    
          ghost/.style={dashed, draw=magenta, postaction={decorate},
        decoration={markings, mark=at position .55 with {\arrow[draw=black]{>}}}
        },  
          antighost/.style={dashed, draw=magenta, postaction={decorate},
        decoration={markings, mark=at position .55 with {\arrow[draw=black]{<}}}
        },  
          mphoton/.style={decorate, decoration={snake}, draw=violet},
            realscalar/.style={draw=black}, 
           mgluon/.style={decorate, draw=blue,
        decoration={coil,amplitude=4pt, segment length=5pt}},
         weylfermion/.style={draw=orange, postaction={decorate},
        decoration={markings,mark=at position .55 with {\arrow[draw=black]{>}}}},
         weylfermionbar/.style={draw=orange, postaction={decorate},
        decoration={markings,mark=at position .55 with {\arrow[draw=black]{<}}}}, 
   	wboson/.style={draw=blue,decorate, decoration={snake,amplitude=4pt}, draw},  
    zboson/.style={draw=violet, decorate, decoration={snake}, draw},   
    lepton/.style={draw=black, postaction={decorate},
        decoration={markings,mark=at position .55 with {\arrow[draw=black]{>}}}},
    leptonbar/.style={draw=black, postaction={decorate},
        decoration={markings,mark=at position .55 with {\arrow[draw=black]{<}}}}, 
        graviton/.style={draw=blue,decorate, decoration={snake,amplitude=4pt}, draw},  
        gravitino/.style={draw=red, postaction={decorate},
        decoration={snake, markings, mark=at position .55 with {\arrow[draw=black]{>}}}},
    gravitinobar/.style={draw=red, postaction={decorate},
        decoration={snake, markings,mark=at position .55 with {\arrow[draw=black]{<}}} },    
}

\allowdisplaybreaks
\begin{document}

\title{Four-Point Yang–Mills Correlators in AdS$_4$ from All-Line Recursion: All-Plus, Single-Minus and MHV}

\author[a]{Dhruv Pathak\footnote{Email: pathakdhruv9786@gmail.com\\ ORCID: 0000-0001-8129-0473}}

\affiliation[a]{Independent Researcher, Delhi, India}
%\email[a]{pathakdhruv9786@gmail.com}

\abstract{
We review Raju's all-line recursion relations for Yang-Mills theory in AdS${}_4$ and evaluate four-point correlators for the All-plus, Single-minus, and MHV helicity configurations. We point out the relevant deformation for the legs needed to obtain the correlators. In particular, we explicitly derive the expressions for all the residues contributing to all three correlators. We also perform an explicit numerical check with the results computed by Witten diagram computations and find perfect agreement with our results. Finally, we conclude by discussing some applications of these recursions for obtaining higher-point functions. 
}

\maketitle

\newpage

%%%%%%%%%%%%%%%%%%%%%%%%%%
\section{Introduction}
\label{sec:introduction}
%%%%%%%%%%%%%%%%%%%%%%%%%%

Over the past two decades the field of scattering amplitude has seen various formalisms to perform computations by analysing its analytic structure: twistor-string methods \cite{Witten:2003nn}, the CSW
expansion \cite{Cachazo:2004kj}, generalised unitarity
\cite{Bern:1994zx,Bern:1994cg,Britto:2004nc}, and above all the
Britto--Cachazo--Feng--Witten (BCFW) recursion relations
\cite{Britto:2004ap,Britto:2005fq}, which reconstruct a tree amplitude from the factorisation channels of lower-point on-shell amplitudes by deforming external momenta into the complex plane. The Risager variant \cite{Risager:2005vk}, in which every shifted momentum is displaced along its own polarisation vector, trades more factorisation terms for weaker conditions on the external data. In parallel, amplitudes have been reorganised in terms of algebraic and geometric structures: color/kinematics duality and the double copy \cite{Bern:2008qj,Kawai:1985xq}, the scattering equations \cite{Cachazo:2013hca}, and positive geometries such as the amplituhedron \cite{Arkani-Hamed:2013jha} and the associahedron
\cite{Arkani-Hamed:2017mur}, in which locality and unitarity are outputs rather than inputs.

The corresponding technology for AdS is less developed. By the AdS/CFT correspondence \cite{Maldacena:1997re,Gubser:1998bc,Witten:1998qj}, Witten diagrams give a conceptually complete prescription for boundary correlators, but in practice a Witten-diagram computation is a Feynman-diagram computation with the additional burden of the radial integrals. Working in Mellin space \cite{Penedones:2010ue,Fitzpatrick:2011ia,Paulos:2011ie} or, for
$\mathrm{AdS}_{4}$, in boundary momentum space \cite{Maldacena:2011nz,Albayrak:2019asr} disposes of the integrals but not the vertices, and for spinning external states --- the conserved currents and stress tensors dual to gluons and gravitons --- the proliferation of tensor structures compounds the problem.

On-shell methods sidestep exactly this: they build higher-point correlators out of lower-point on-shell data, and never expand the bulk Lagrangian at all. In AdS, however, such tools are scarce. The main line of work is a series of papers by Raju \cite{Raju:2010by,Raju:2011mp,Raju:2012zr,Raju:2012zs}, generalising BCFW to correlators of conserved currents and of the stress tensor. The idea is that the external momenta can be deformed by a complex parameter $w$
in a way that preserves both momentum conservation and each norm $|\bm{k}_{i}|$. The radial wavefunctions are then blind to the deformation, so that at fixed radial momentum $p$, the integrand is a rational function of $w$ whose poles occur only where a bulk-to-bulk propagator goes on shell. At such a pole, the numerator of the propagator degenerates into a sum over normalisable modes, and the residue factorises into lower-point \emph{transition amplitudes}: correlators in which one bulk-to-boundary propagator has been replaced by a normalisable mode \cite{Balasubramanian:1999ri}. The integrals over $p$ are performed afterwards; in $d=3$ they too reduce to residues at poles of two
kinds: those of the constituent transition amplitudes and one from
the propagator.

Despite being, to our knowledge, essentially the only genuinely on-shell
recursion for spinning fields in AdS, this technology has seen very little
subsequent use; we are aware of one application, to the all-plus four-point
graviton correlator \cite{Albayrak:2023jzl}. The checks performed on this technology are moreover dominated by the flat-space limit, in which the coefficient of the leading singularity at vanishing total energy, $E=\sum_{i}|\bm{k}_{i}|\to0$, is matched to the known $(d+1)$-dimensional amplitude \cite{Maldacena:2011nz,Raju:2012zr}. Such a check probes only the leading singularity and is blind to the intrinsically AdS part of the answer.

We are not aware of any literature in which the output of this AdS recursion has been compared against an independent Witten diagram computation away from the flat-space limit. This paper supplies this verification and uses the recursion to produce new results. We review the all-line recursion relations of \cite{Raju:2012zr,Raju:2012zs} for current correlators in Yang--Mills theory in $\mathrm{AdS}_{4}$. We give the essential ingredients to compute these correlators: the three-dimensional spinor-helicity conventions, the three-point transition amplitudes, the radial-momentum integral $R^{\rm YM}$ which serves as the leading factor in the cubic vertex, and the pole locations. We then compute the four-point color-ordered current correlator in closed analytic form for three helicity assignments: (i) All-plus case ($T(++++)$), the Single-minus case ($T(-+++)$), and the MHV case ($T(-+-+)$). The corresponding $p$-integrals can be computed by performing contour integration and obtaining residues for relevant poles. Each color-ordered partition contributes three poles when we close the contour in the upper-half $p$-plane. We give the derivations of obtaining the residues, highlighting the non-trivial steps encountered in doing so. We check these expressions numerically, at generic kinematics rather than in the flat-space limit, against the $\mathrm{AdS}_{4}$ Yang--Mills amplitudes of \cite{ALM:2021}, obtained from Witten diagram computations, and get a perfect match. One convention-dependent point affects the comparison: the cubic vertex of \cite{ALM:2021} differs by a factor of $-i$ from that of \cite{Raju:2012zs}, and since the exchange contributions are quadratic in the vertex this produces an overall relative sign, which we track explicitly. 

Our choice to carry all three helicity configurations in parallel is not merely a matter of completeness. In flat-space the All-plus and Single-minus tree amplitudes vanish identically, and MHV is the first non-vanishing
configuration but in AdS all these configurations exist. In fact, for computing higher point correlators in AdS via recursion relations, we would need four-point transition amplitudes of these very helicity configurations.

\paragraph{Outline and conventions.} 
Section~\ref{sec:alllinerev} reviews the all-line recursion relations given by \cite{Raju:2012zs}, which we use in this work. 
Sections~\ref{sec:allplus1}, \ref{sec:singlmin1}, and \ref{sec:mhv1} work out all the residue contributions to All-plus, Single-minus and MHV correlators, respectively. Section~\ref{sec:check} presents the numerical
comparison with the Witten diagram computation results based on work by \cite{ALM:2021}. We conclude in Section~\ref{sec:conclusions}. Appendix~\ref{sec:spinform} sets up the spinor-helicity formalism as well as conventions we use in this work. Appendix~\ref{sec:3point} gives three-point transition amplitudes used in our work. In Appendix \ref{app:schan} we discuss in detail some of the intricacies in the derivations of residues. Appendix \ref{sec:witten} gives the expressions of the four-point correlators obtained by \cite{ALM:2021} using Witten diagram computations. We use the mostly-plus signature $\eta_{\mu\nu}=\mathrm{diag}(-1,+1,+1,+1)$. Boldface denotes boundary vectors; all correlators are color-ordered partial amplitudes.

\section{Correlator Computation using All-line Recursion Relations}
\label{sec:bcfw}
%%%%%%%%%%%%%%%%%%%%%%%%%%
\subsection{Review of the All-line Recursion Relations}
\label{sec:alllinerev}
%%%%%%%%%%%%%%%%%%%%%%%%%%

We give here the recursion relations for conserved current correlators given by \cite{Raju:2012zs} for completeness. These include all-line deformations of spinors, unlike the standard BCFW two-line deformations \cite{Raju:2011mp}. However, these relations work well for three-dimensional systems, unlike the standard two-line deformations, in the sense that the boundary term always vanishes at infinity.

\begin{equation}
\label{recurs1}
\begin{split}
&T(h_1, \bm{k_1}(w), \ldots h_4, \bm{k_4}(w)) =  \int_{-\infty}^{\infty} \left[ \sum_{\pi} { I}_{\pi}(w,p) + { B}(w,p) \right] d p,  \\ 
&{ I}_{\pi}(w,p) = {p \over 4} \sum_{h^{\text{int}}, \pm} {-i {\cal T}^2 \over p^2 + (\bm{k_{\pi_1}}(w) + \bm{k_{\pi_2}}(w))^2}  {w-w^{\mp}(p) \over w^{\pm}(p) - w^{\mp}(p)},
\\
&{\cal T}^2 \equiv  T({h_{\pi_1}}, \bm{k_{\pi_1}}(p), {h_{\pi_2}}, \bm{k_{\pi_2}}(p), {h_{\text{int}}}, \bm{k_{\text{int}}}) T({-h_{\text{int}}}, -\bm{k_{\text{int}}},{h_{\pi_3}}, \bm{k_{\pi_3}}(p),  {h_{\pi_4}}, \bm{k_{\pi_4}}(p)).
\end{split}
\end{equation}

Here, $\pi$ in the subscript is the kind of partition, e.g., (12)(34) and (14)(32). $h$ is the helicity of the external particle (gluon in our case). $B$ is the boundary term which causes $p$ integral to converge by fixing the behaviour of the integrand at large $p$ and large $w$. It is a polynomial in $w$ with rational coefficients and so, doesn't contribute to any pole. 

We need the undeformed correlator, which we get by setting $w=0$,
\begin{equation}
\label{recursw0}
{ I}_{\pi}(0,p)
=\frac{p}{4}
\sum_{h_{\rm int},\pm}
\frac{i{\cal T}^2}
{p^2+(\bm{k}_{\pi_1}+\bm{k}_{\pi_2})^2}
\frac{w^{\mp}(p)}
{w^{\pm}(p)-w^{\mp}(p)},
\end{equation}

and the four-point function is obtained from the residues as,
\begin{equation}
\label{Teres}
T(h_1,\bm{k_1},\ldots,h_4,\bm{k_4})
=
2\pi i \sum_{\pi} \sum_{p_0\in{\cal P}_{\pi}}
\underset{p=p_0}{\operatorname{Res}}
\left[{I}_{\pi}(0,p)\right].
\end{equation}

As each partition contributes three poles when we close the contour in the upper-half $p$-plane, the poles for the (12)(34) partition (denoted by $s$ subscript) are,
\begin{equation}
\mathcal{P}_{s}
=
\Big\{\,
i\!\left(|\bm{k}_{1}|+|\bm{k}_{2}|\right),\;\;
i\!\left(|\bm{k}_{3}|+|\bm{k}_{4}|\right),\;\;
i\left|\bm{k}_{1}+\bm{k}_{2}\right|
\,\Big\}.
\label{eq:intro-poles}
\end{equation}

The recursion relations work for the color-ordered amplitudes. Hence, the total amplitude is the sum of the amplitudes obtained from the (12)(34) partition and the (14)(32) partition. Once we obtain the residues from the (12)(34) partition, we just need to interchange $2\leftrightarrow4$ to obtain the expressions for the residues from the (14)(32) partition. Sum of all the residues from the (12)(34) partition and the (14)(32) partition gives us the contribution to the four-point correlator.

%%%%%%%%%%%%%%%%%%%%%%%%%%
\subsection{All Plus case ($T(++++)$) Computation}
\label{sec:allplus1}

\begin{figure}[h]
    \centering
    \includegraphics[width=0.5\linewidth]{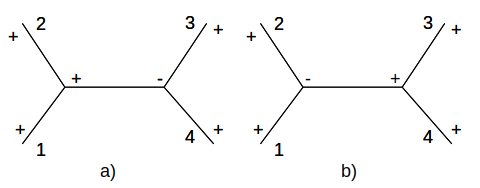}
    \caption{All-plus case}
    \label{fig:allplus}
\end{figure}

Deformations for the all-plus case are,
\begin{align}
    \lambda_{i}(w) &= \lambda_{i} + \beta_{i} w \bar{\lambda}_{i}; \, \bar{\lambda}_{i} (w) = \bar{\lambda}_{i}, \label{eq:allpdeform}
\end{align}
where the ratios of $\beta_i$ are fixed from the momentum conservation equation and appropriate contractions. For the all-plus case, momentum conservation gives,
\begin{align}
  &\lambda_1^{\alpha} (w) \bar{\lambda}_1^{\dot{\alpha}}  + \lambda_2^{\alpha} (w) \bar{\lambda}_2^{\dot{\alpha}}  + \lambda_3^{\alpha} (w) \bar{\lambda}_3^{\dot{\alpha}} + \lambda_4^{\alpha} (w) \bar{\lambda}_4^{\dot{\alpha}} = -E \, e^{\alpha\,\dot{\alpha}},  \\
  \implies &  \beta_1 \bar{\lambda}_1^{\alpha} \bar{\lambda}_1^{\dot{\alpha}} + \beta_2 \bar{\lambda}_2^{\alpha} \bar{\lambda}_2^{\dot{\alpha}}  + \beta_3  \bar{\lambda}_3^{\alpha} \bar{\lambda}_3^{\dot{\alpha}} + \beta_4 \bar{\lambda}_4^{\alpha} \bar{\lambda}_4^{\dot{\alpha}} = 0 .
\end{align}

So, we have,
\begin{align}
    \frac{\beta_2}{\beta_1} = -\frac{\langle \bar{\lambda}_4 \bar{\lambda}_1 \rangle \, \langle  \bar{\lambda}_3 \bar{\lambda}_1 \rangle}{\langle \bar{\lambda}_4 \bar{\lambda}_2 \rangle \, \langle  \bar{\lambda}_3 \bar{\lambda}_2 \rangle}, \, \frac{\beta_3}{\beta_1} = -\frac{\langle \bar{\lambda}_2 \bar{\lambda}_1 \rangle \, \langle  \bar{\lambda}_4 \bar{\lambda}_1 \rangle}{\langle \bar{\lambda}_2 \bar{\lambda}_3 \rangle \, \langle  \bar{\lambda}_4 \bar{\lambda}_3 \rangle},\, \frac{\beta_4}{\beta_1} = -\frac{\langle \bar{\lambda}_2 \bar{\lambda}_1 \rangle \, \langle  \bar{\lambda}_3 \bar{\lambda}_1 \rangle}{\langle \bar{\lambda}_2 \bar{\lambda}_4 \rangle \, \langle  \bar{\lambda}_3 \bar{\lambda}_4 \rangle} .
\end{align}
We remind the reader that the energy is not conserved. 

By plugging in the expressions for the corresponding three-point functions given in appendix \ref{sec:3point} into \eqref{recursw0}, we obtain, 
\begin{align}
  \hskip-1.cm  I_s(0) 
    &=\sum_{\pm} \frac{1}{16 \pi |\bm{k_1}||\bm{k_2}| |\bm{k_3}||\bm{k_4}|} \frac{-1}{p^2 + (\bm{k_1} + \bm{k_2})^2} \nonumber \\
    &\hskip2cm \times \left[ \frac{(|\bm{k_4}| -ip - |\bm{k_3}|)(-ip + |\bm{k_3}| -|\bm{k_4}|)}{(|\bm{k_1}|+|\bm{k_2}| - ip)((|\bm{k_3}|+|\bm{k_4}| - ip))}\frac{\langle \bar{\lambda}_1 \bar{\lambda}_2  \rangle  \langle \bar{\lambda}_2  \bar{\lambda}_{int}\rangle \langle \bar{\lambda}_{int} \bar{\lambda}_1 \rangle \langle \bar{\lambda}_3 \bar{\lambda}_4  \rangle^3}{\langle \bar{\lambda}_4  \bar{\lambda}_{int} \rangle \langle \bar{\lambda}_{int} \bar{\lambda}_3 \rangle}\right. \nonumber \\
    &\hskip1cm \left.+ \frac{(|\bm{k_2}| +ip - |\bm{k_1}|)(ip + |\bm{k_1}| -|\bm{k_2}|)}{(|\bm{k_1}|+|\bm{k_2}| + ip)(|\bm{k_3}|+|\bm{k_4}| + ip)} \frac{\langle \bar{\lambda}_1 \bar{\lambda}_2  \rangle^3 \langle \bar{\lambda}_3 \bar{\lambda}_4  \rangle  \langle \bar{\lambda}_4  \bar{\lambda}_{int}\rangle \langle \bar{\lambda}_{int} \bar{\lambda}_3 \rangle}{\langle \bar{\lambda}_2  \bar{\lambda}_{int} \rangle \langle \bar{\lambda}_{int} \bar{\lambda}_1 \rangle} \right] \frac{w^{\mp}}{w^{\pm} - w^{\mp}} .
    \label{eq:allpschan}
\end{align}

Here, the subscript $s$ denotes the (12)(34) partition and `$(0)$' denotes the undeformed integrand at $w=0$.

To obtain the final answer $T(++++)$, we must perform the integral over the internal radial momentum $p$. In AdS$_4$, the relevant bulk-to-boundary and bulk-to-bulk propagators reduce to elementary functions, owing to the half-integer orders of the Bessel functions. Consequently, the integrand appearing in \eqref{recurs1} is a rational function of $p$ (and is even in $p$), allowing the $p$ integral to be evaluated by contour integration and residues. There are 3 distinct pairs of poles in the $p$ plane, and we choose to close the contour in the upper half-plane, 
$p=i(|\bm{k_1}|+|\bm{k_2}|),\,p=i(|\bm{k_3}|+|\bm{k_4}|),\,p= i|\bm{k_1} + \bm{k_2}|$. In each case $w$ is determined accordingly; see Appendix C for the derivation. We state the final result of the residue computation at each of the poles below. 
\begin{align}
&2\pi i\underset{p=i(|\bm{k_1}|+|\bm{k_2}|)}{\operatorname{Res}}\,I_s^{(0)}
={}
\frac{\langle\bar{\lambda}_3\bar{\lambda}_4\rangle}
{8|\bm{k_1}||\bm{k_2}| |\bm{k_3}||\bm{k_4}|(|\bm{k_3}| + |\bm{k_4}| - |\bm{k_1}| - |\bm{k_2}|)\langle\lambda_1\lambda_2\rangle}
\nonumber\\
&\times
\Bigg\{
4|\bm{k_1}||\bm{k_2}|
\left[
\langle\bar{\lambda}_1\bar{\lambda}_4\rangle
\langle\bar{\lambda}_2\bar{\lambda}_3\rangle
+
\langle\bar{\lambda}_2\bar{\lambda}_4\rangle
\langle\bar{\lambda}_1\bar{\lambda}_3\rangle
\right]
-2|\bm{k_2}|
\langle\bar{\lambda}_2\bar{\lambda}_4\rangle
\langle\bar{\lambda}_2\bar{\lambda}_3\rangle
\langle\lambda_2\bar{\lambda}_1\rangle -2|\bm{k_1}|
\langle\bar{\lambda}_1\bar{\lambda}_4\rangle
\langle\bar{\lambda}_1\bar{\lambda}_3\rangle
\langle\lambda_1\bar{\lambda}_2\rangle
\Bigg\}. \label{pik12res1}
\end{align}

The residue at the pole $p=i(|\bm{k_3}|+|\bm{k_4}|)$, is obtained by just interchanging $(1,2) \leftrightarrow (3,4)$ in the above expression for the residue at $p=i(|\bm{k_1}|+|\bm{k_2}|)$ pole. We can also see from eq. \eqref{pik12res1} that there is no total energy ($E$) pole.

For the residue at the pole $p= i|\bm{k_1} + \bm{k_2}|$, we get a contribution from just one $w$, where $w = 0$. This formalism is explained in section 5.1 of \cite{Raju:2012zs}. The corresponding residue is, 
\begin{align}\label{pinKsres2}
  \hskip-1cm &2\pi i\underset{p=i(|\bm{k_1}+\bm{k_2}|)}{\operatorname{Res}}\,I_s^{(0)} \nonumber \\
    &= \frac{1}{16|\bm{k_1}||\bm{k_2}| |\bm{k_3}||\bm{k_4}|} \frac{1}{|\bm{k_1} + \bm{k_2}|}\left[ \frac{(|\bm{k_4}| + |\bm{k_1} + \bm{k_2}| - |\bm{k_3}|)(|\bm{k_1} + \bm{k_2}|+ |\bm{k_3}| -|\bm{k_4}|)}{(|\bm{k_1}|+|\bm{k_2}| +|\bm{k_1} + \bm{k_2}|)((|\bm{k_3}|+|\bm{k_4}| + |\bm{k_1} + \bm{k_2}|))} \langle \bar{\lambda}_1 \bar{\lambda}_2  \rangle \langle \bar{\lambda}_3 \bar{\lambda}_4 \rangle^3 \right. \nonumber \\
    &\hskip1cm \left. \frac{(E^{34}_{-s}\langle \lambda_3 \bar{\lambda}_2  \rangle  +  \langle \lambda_3 \lambda_4  \rangle \langle \bar{\lambda}_2 \bar{\lambda}_4  \rangle )(E^{34}_{-s}\langle \lambda_4 \bar{\lambda}_1  \rangle  +  \langle \lambda_4 \lambda_3  \rangle \langle \bar{\lambda}_1 \bar{\lambda}_3  \rangle )}{(E^{34}_{-s})^2 \langle \bar{\lambda}_4 \lambda_3 \rangle \langle \bar{\lambda}_3 \lambda_4 \rangle} \right. \nonumber \\
    &\hskip1cm \left. + \frac{(|\bm{k_2}| -|\bm{k_1} + \bm{k_2}| - |\bm{k_1}|)(-|\bm{k_1} + \bm{k_2}| + |\bm{k_1}| -|\bm{k_2}|)}{(|\bm{k_1}|+|\bm{k_2}| -|\bm{k_1} + \bm{k_2}|)(|\bm{k_3}|+|\bm{k_4}|- |\bm{k_1} + \bm{k_2}|)} \right. \nonumber \\
    & \hskip1.5cm \left.  \langle \bar{\lambda}_1 \bar{\lambda}_2  \rangle^3 \langle \bar{\lambda}_3 \bar{\lambda}_4 \rangle \frac{(E^{12}_{s}\langle \lambda_1 \bar{\lambda}_4  \rangle  +  \langle \lambda_1 \lambda_2  \rangle \langle \bar{\lambda}_4 \bar{\lambda}_2  \rangle )(E^{12}_{s}\langle \lambda_2 \bar{\lambda}_3  \rangle  +  \langle \lambda_2 \lambda_1  \rangle \langle \bar{\lambda}_3 \bar{\lambda}_1  \rangle )}{(E^{12}_{s})^2 \langle \bar{\lambda}_2 \lambda_1 \rangle \langle \bar{\lambda}_1 \lambda_2 \rangle} \right].
\end{align}

$E^{12}_{s}$ and $E^{34}_{-s}$ are defined in eq. (\ref{eq:Edefs})

To get the full answer, we also need to include the contribution from the other partition, which is obtained by interchanging 2 and 4 in the previous expression.

Hence, the final answer is obtained as follows
\begin{eqn}
T(++++) = \text{\eqref{pik12res1} + \eqref{pik12res1}$|_{(1,2)\leftrightarrow (3,4)}$ + \eqref{pinKsres2} + $(2\leftrightarrow 4)$} .
\end{eqn}

\subsection{Single-minus case ($T(- + + + )$) Computation}
\label{sec:singlmin1}

\begin{figure}[h]
    \centering
    \includegraphics[width=0.5\linewidth]{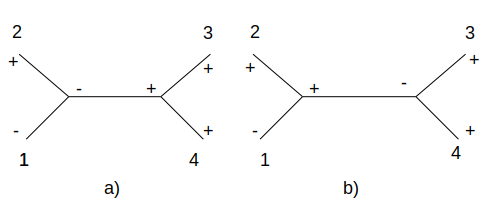}
    \caption{Single-minus Case}
    \label{fig:NMHV}
\end{figure}

The deformations in this case differ from those of the all-plus case. As explained in \cite{Raju:2012zs}, the $\bar\lambda$ of the negative helicity gluon is deformed in the following manner,

For $i=1$,
\begin{align}
    \lambda_{1}(w) = \lambda_{1} ; \, \bar{\lambda}_{1} (w) = \bar{\lambda}_{1} + \beta_{1} w \lambda_{1}. \label{eq:nmhvdef1}
\end{align}
The positive helicity gluons follow a pattern similar to the all-plus case,

For $i=2,3, \& \,4$,
\begin{align}
    \lambda_{i}(w) = \lambda_{i} + \beta_{i} w \bar{\lambda}_{i}; \, \bar{\lambda}_{i} (w) = \bar{\lambda}_{i} .\label{eq:nmhvdef2}
\end{align}
The procedure for determining $\beta_i$ is exactly the same as the all-plus case, and we obtain,
\begin{align}
    \frac{\beta_2}{\beta_1} = -\frac{\langle \bar{\lambda}_4 \lambda_1 \rangle \, \langle  \bar{\lambda}_3 \lambda_1 \rangle}{\langle \bar{\lambda}_4 \bar{\lambda}_2 \rangle \, \langle  \bar{\lambda}_3 \bar{\lambda}_2 \rangle},\, \frac{\beta_3}{\beta_1} = -\frac{\langle \bar{\lambda}_2 \lambda_1 \rangle \, \langle  \bar{\lambda}_4 \lambda_1 \rangle}{\langle \bar{\lambda}_2 \bar{\lambda}_3 \rangle \, \langle  \bar{\lambda}_4 \bar{\lambda}_3 \rangle},\, \frac{\beta_4}{\beta_1} = -\frac{\langle \bar{\lambda}_2 \lambda_1 \rangle \, \langle  \bar{\lambda}_3 \lambda_1 \rangle}{\langle \bar{\lambda}_2 \bar{\lambda}_4 \rangle \, \langle  \bar{\lambda}_3 \bar{\lambda}_4 \rangle}. \label{eq:betanmhv}
\end{align}

By plugging in the expressions for the corresponding three-point functions given in appendix \ref{sec:3point} into \eqref{recurs1} we obtain, 

\begin{align}
  \hskip-1.5cm  I_s(0) 
    &=\sum_{\pm} \frac{1}{16 \pi |\bm{k_1}||\bm{k_2}| |\bm{k_3}||\bm{k_4}|} \frac{-1}{p^2 + (\bm{k_1} + \bm{k_2})^2} \nonumber \\
    &\hskip2cm \times \left[  \frac{(|\bm{k_2}| +ip - |\bm{k_1}|)(ip + |\bm{k_1}| -|\bm{k_2}|)}{(|\bm{k_1}|+|\bm{k_2}| + ip)(|\bm{k_3}|+|\bm{k_4}| + ip)} \frac{\langle \lambda_{int} \lambda_1  \rangle^3 \langle \bar{\lambda}_3 \bar{\lambda}_4  \rangle  \langle \bar{\lambda}_4  \bar{\lambda}_{int}\rangle \langle \bar{\lambda}_{int} \bar{\lambda}_3 \rangle}{\langle \lambda_1  \lambda_2(w) \rangle \langle \lambda_2 (w) \lambda_{int} \rangle}\right. \nonumber \\
    & \hskip2cm \left.+ \frac{(|\bm{k_2}| +ip - |\bm{k_1}|)(ip + |\bm{k_1}| -|\bm{k_2}|)(|\bm{k_4}| -ip - |\bm{k_3}|)(-ip + |\bm{k_3}| -|\bm{k_4}|)}{(|\bm{k_1}|+|\bm{k_2}| + ip)((|\bm{k_3}|+|\bm{k_4}| - ip))} \right. \nonumber \\
    &\hskip2cm \left. \frac{\langle \bar{\lambda}_2 \bar{\lambda}_{int}  \rangle^3 \, \langle \bar{\lambda}_3 \bar{\lambda}_4  \rangle^3}{\langle \bar{\lambda}_{int}  \bar{\lambda}_1 (w) \rangle \langle \bar{\lambda}_1 (w) \bar{\lambda}_2 \rangle\, \langle \bar{\lambda}_4  \bar{\lambda}_{int} \rangle \langle \bar{\lambda}_{int} \bar{\lambda}_3 \rangle} \right] \frac{w^{\mp}}{w^{\pm} - w^{\mp}} . \label{eq:nmhvschan}
\end{align}

The final answer through the recursion relation for this case is obtained by a similar set of manipulations as the all-plus case. The final answer is again obtained as a sum over six terms, with each representing the residue of the $p$ integral on the upper half-plane. For the $(12)(34)$ partition we have the contribution at $p = i (|\bm k_1| + |\bm k_2|), p = i (|\bm k_3| + |\bm k_4|)$ and $p = i |\bm k_1 + \bm k_2|$. Similarly for the $(14)(32)$ partition. We give the residues of each of the relevant poles below.\\

\hskip-0.8cm\textbf{(A) For $p=i(|\bm{k_1}|+|\bm{k_2}|)$ pole}

For this case, we have,

\begin{align}
 \langle \lambda_{1} \lambda_{2}(w) \rangle \langle \bar{\lambda}_{1} (w)\bar{\lambda}_{2}\rangle = 0 .\label{eq:nmhv12w+w-}
\end{align}
Hence, we have two solutions for $w$: a) $\langle \lambda_{1} \lambda_{2}(w^+) \rangle=0$ and b) $\langle \bar{\lambda}_{1} (w^-) \bar{\lambda}_{2}\rangle=0$.

\hskip-0.5cm\textbf{i) When $w=w^+$:}

The non-vanishing contribution to the residue for the $w=w^+$ case comes from diagram (b) in Fig. \ref{fig:NMHV}, which is given by,
\begin{align}
\hskip-1cm &2\pi i  \underset{\substack{p=i(|\bm{k_1}|+|\bm{k_2}|)\\ w=w^+} }{\operatorname{Res}}\,I_s^{(0)} \nonumber \\
    &=  \frac{-1}{8 |\bm{k_1}||\bm{k_2}| |\bm{k_3}||\bm{k_4}|}\frac{2|\bm{k_1}|\, 2|\bm{k_2}|}{\langle \lambda_1 \lambda_2 \rangle \langle \bar{\lambda}_1 (w^+)  \bar{\lambda}_2 \rangle} \left[ \frac{(E - 2|\bm{k_3}|)(E -2|\bm{k_4}|)}{E} \right. \nonumber \\
    &\hskip0.7cm \left.\times \frac{\left(E  \langle \lambda_3 (w^+)  \bar{\lambda}_2 \rangle + \langle \lambda_3 (w^+) \lambda_4 (w^+) \rangle  \langle \bar{\lambda}_2  \bar{\lambda}_4 \rangle\right)^2 \left(E  \langle \lambda_4 (w^+)  \bar{\lambda}_2 \rangle + \langle \lambda_4 (w^+) \lambda_3 (w^+) \rangle  \langle \bar{\lambda}_2  \bar{\lambda}_3 \rangle \right)}{\left(E  \langle \lambda_3 (w^+)  \bar{\lambda}_1 (w^+) \rangle + \langle \lambda_3 (w^+) \lambda_4 (w^+) \rangle  \langle \bar{\lambda}_1 (w^+) \bar{\lambda}_4 \rangle \right) \langle \bar{\lambda}_1 (w^+) \bar{\lambda}_2  \rangle}\right. \nonumber \\
    &\hskip1cm \left. \times \frac{\langle \bar{\lambda}_3 \bar{\lambda}_4  \rangle^3 }{E^2   \langle \lambda_3 (w^+)  \bar{\lambda}_4 \rangle  \langle \lambda_4 (w^+)  \bar{\lambda}_3 \rangle} \right].\label{eq:nmhvw+k1k2}
\end{align}
The values of $\bar{\lambda}_1 (w^+)$, $\lambda_2 (w^+)$, $\lambda_3 (w^+)$, and $\lambda_4 (w^+)$ can be obtained by expanding them in terms of $\beta_i w^+$, which can be written as $\frac{\beta_i}{\beta_2}\,\beta_2 w^+$. We can then substitute values of $\frac{\beta_i}{\beta_2}$ from eq. \eqref{eq:betanmhv} and $\beta_2 w^+$ from eq. \eqref{eq:b2w+}.

The total energy $E$ appearing in the denominator in eqs. \eqref{eq:nmhvw+k1k2} is a spurious pole. We see that there are 3 powers of $E$ in the denominator. After using eq. \eqref{eq:allspinorsum}, left hand side of eq. \eqref{eq:nmhvw+k1k2} can be reduced to
\begin{align}
   \frac{2|\bm{k_1}|\, 2|\bm{k_2}|(E - 2|\bm{k_3}|)(E -2|\bm{k_4}|)}{8 |\bm{k_1}||\bm{k_2}| |\bm{k_3}||\bm{k_4}|} \frac{\langle \lambda_1 \bar\lambda_4 \rangle^2 \langle \lambda_1 \bar\lambda_3 \rangle }{\langle \lambda_3 (w^+) \lambda_2 (w^+) \rangle  \langle \lambda_3 (w^+)  \bar{\lambda}_4 \rangle  \langle \lambda_4 (w^+)  \bar{\lambda}_3 \rangle\langle \lambda_1 \lambda_2 \rangle},
\end{align}
which is devoid of $E$ pole.

\iffalse
Two of them can be combined with $\langle \lambda_3 (w^+)  \bar{\lambda}_4 \rangle $ and  $ \langle \lambda_4 (w^+)  \bar{\lambda}_3 \rangle$ as,
\begin{align}
    &E  \langle \lambda_4 (w^+)  \bar{\lambda}_3 \rangle = -(\langle \lambda_4 (w^+)  \lambda_1  \rangle \langle \bar\lambda_3 \bar{\lambda}_1 (w^+) \rangle + \langle \lambda_4 (w^+)  \lambda_2 (w^+) \rangle \langle \bar\lambda_3  \bar{\lambda}_2 \rangle ) \\
   & E  \langle \lambda_3 (w^+)  \bar{\lambda}_4 \rangle = -(\langle \lambda_3 (w^+)  \lambda_1 \rangle \langle \bar\lambda_4  \bar{\lambda}_1 (w^+) \rangle + \langle \lambda_3 (w^+)  \lambda_2 (w^+)\rangle \langle \bar\lambda_4   \bar{\lambda}_2 \rangle ) 
\end{align}
\fi

\hskip-0.5cm\textbf{ii) When $w=w^-$:}

The non-vanishing contribution to the residue for the $w=w^-$ case comes from diagram (a) in Fig. \ref{fig:NMHV}, which is given by,
\begin{align}
\hskip-1cm &2\pi i \underset{\substack{p=i(|\bm{k_1}|+|\bm{k_2}|)\\ w=w^-} }{\operatorname{Res}}\,I_s^{(0)} \nonumber \\
&=   \frac{-1}{8 |\bm{k_1}||\bm{k_2}| |\bm{k_3}||\bm{k_4}|}\frac{2|\bm{k_1}| \,2|\bm{k_2}|}{\langle \lambda_1 \lambda_2 (w^-) \rangle \langle \bar{\lambda}_1  \bar{\lambda}_2 \rangle} \left[  \frac{1}{(|\bm{k_3}|+|\bm{k_4}| -|\bm{k_1}|-|\bm{k_2}|)} \right. \nonumber \\ 
&\hskip0.3cm\left. \times\frac{\left(E \langle \lambda_1  \bar{\lambda}_3 \rangle + \langle \lambda_1 \lambda_4 (w^-) \rangle  \langle \bar{\lambda}_3 \bar{\lambda}_4 \rangle\right)^2 \left(E  \langle \lambda_1  \bar{\lambda}_4 \rangle + \langle \lambda_1 \lambda_3 (w^-) \rangle  \langle \bar{\lambda}_4 \bar{\lambda}_3 \rangle \right)  \langle \bar{\lambda}_3 \bar{\lambda}_4  \rangle }{\langle \lambda_1 \lambda_2 (w^-) \rangle  \left(E \langle \lambda_2 (w^-)  \bar{\lambda}_3 \rangle + \langle \lambda_2 (w^-) \lambda_4 (w^-) \rangle  \langle \bar{\lambda}_3 \bar{\lambda}_4 \rangle \right)}\right] .\label{eq:nmhvw-k1k2}
\end{align}
The values of $\bar{\lambda}_1 (w^-)$, $\lambda_2 (w^-)$, $\lambda_3 (w^-)$, and $\lambda_4 (w^-)$ can be obtained by expanding them in terms of $\beta_i w^-$, which can be written as $\frac{\beta_i}{\beta_1}\,\beta_1 w^-$. We can then substitute values of $\frac{\beta_i}{\beta_1}$ from eq. \eqref{eq:betanmhv} and $\beta_1 w^-$ from eq. \eqref{eq:b1w-}.

Note that in eq. \eqref{eq:nmhvw-k1k2}, as $E\rightarrow0$, there is no singularity, so there is no total energy pole.

\vskip0.5cm

\hskip-0.8cm\textbf{(B) For $p=i(|\bm{k_3}|+|\bm{k_4}|)$ pole}

\begin{align}
    &\hskip-1cm 2 \pi i \underset{\substack{p=i (|\bm{k_3}|+|\bm{k_4}|)}}{\operatorname{Res}}\,I_s^{(0)} \nonumber \\
     &= \frac{-1}{8  |\bm{k_1}||\bm{k_2}| |\bm{k_3}||\bm{k_4}|} \frac{(|\bm{k_2}|- |\bm{k_3}|-|\bm{k_4}| - |\bm{k_1}|)( |\bm{k_1}| - |\bm{k_3}|-|\bm{k_4}| -|\bm{k_2}|)(|\bm{k_1}|+|\bm{k_2}| -|\bm{k_3}|-|\bm{k_4}|)}{\langle \lambda_3 \lambda_4 \rangle  (|\bm{k_1}| - |\bm{k_2}|+|\bm{k_3}|+|\bm{k_4}| ) } \nonumber \\
    &\hskip0.5cm \times \langle  \lambda_1 \bar{\lambda}_2 \rangle^3 2 |\bm{k_4}| 2 |\bm{k_3}| 2 |\bm{k_3}| \sum_{\pm}  \left[  \frac{ 1}{ (\langle \bar{\lambda}_3 \lambda_4 \rangle + \beta_4 w \langle \bar{\lambda}_3 \bar{\lambda}_4  \rangle)    \langle  \bar{\lambda}_2 \lambda_3 (w) \rangle^2 \langle \lambda_2 (w) \lambda_1 \rangle }\right] \frac{w^{\mp}}{w^{\pm} - w^{\mp}} \,.  \label{eq:nmhvk3k4a} 
\end{align}
The explicit expression for the summed-over part is derived in Appendix C (see eqs. \ref{eq:nmhvk3k4A} and \ref{eq:f0nmhv}).

\vskip0.5cm

\hskip-0.8cm\textbf{(C) For $p=i(|\bm{k_1}+\bm{k_2}|)$ pole}

\begin{align}
    &\hskip-1cm 2\pi i  \underset{\substack{p=i|\bm{k_1}+\bm{k_2}|} }{\operatorname{Res}}\,I_s^{(0)} \nonumber \\
    &=  \frac{1}{16  |\bm{k_1}||\bm{k_2}| |\bm{k_3}||\bm{k_4}|} \frac{(|\bm{k_2}| +ip - |\bm{k_1}|)(ip + |\bm{k_1}| -|\bm{k_2}|)}{|\bm{k_1}+\bm{k_2}| (|\bm{k_1}|+|\bm{k_2}| + ip) } \nonumber \\
    &\hskip0.5cm \times \left[  \frac{1}{(|\bm{k_3}|+|\bm{k_4}| + ip)} \frac{\left(E^{34}_{-s}  \langle \lambda_1  \bar{\lambda}_3 \rangle + \langle \lambda_1 \lambda_4  \rangle  \langle \bar{\lambda}_3 \bar{\lambda}_4 \rangle\right)^2 \left(E^{34}_{-s}  \langle \lambda_1  \bar{\lambda}_4 \rangle + \langle \lambda_1 \lambda_3 \rangle  \langle \bar{\lambda}_4 \bar{\lambda}_3 \rangle \right)  \langle \bar{\lambda}_3 \bar{\lambda}_4  \rangle }{\langle \lambda_1  \lambda_2 \rangle \left(E^{34}_{-s}  \langle \lambda_2   \bar{\lambda}_3 \rangle + \langle \lambda_2  \lambda_4  \rangle  \langle \bar{\lambda}_3 \bar{\lambda}_4 \rangle \right)} \right. \nonumber \\
    &\hskip0.8cm \left.+ \frac{(|\bm{k_4}| -ip - |\bm{k_3}|)(-ip + |\bm{k_3}| -|\bm{k_4}|)}{(|\bm{k_3}|+|\bm{k_4}| - ip)} \frac{\left(E^{34}_{-s}  \langle \lambda_3  \bar{\lambda}_2 \rangle + \langle \lambda_3  \lambda_4  \rangle  \langle \bar{\lambda}_2  \bar{\lambda}_4 \rangle\right)^2 \left(E^{34}_{-s}  \langle \lambda_4  \bar{\lambda}_2 \rangle + \langle \lambda_4  \lambda_3 \rangle  \langle \bar{\lambda}_2  \bar{\lambda}_3 \rangle \right) }{\left(E^{34}_{-s}  \langle \lambda_3   \bar{\lambda}_1  \rangle + \langle \lambda_3  \lambda_4  \rangle  \langle \bar{\lambda}_1  \bar{\lambda}_4 \rangle \right)  \langle \bar{\lambda}_1  \bar{\lambda}_2 \rangle  (E^{34}_{-s})^2  } \right. \nonumber \\
    &\hskip2cm\left. \times \frac{\langle \bar{\lambda}_3 \bar{\lambda}_4  \rangle^3}{\langle \lambda_3   \bar{\lambda}_4 \rangle   \langle \lambda_4  \bar{\lambda}_3 \rangle}\right] \Big|_{p=i|\bm{k_1}+\bm{k_2}|} . \label{eq:nmhvk12}
\end{align}
The expression for $E^{34}_{-s}$ is given in eq. \eqref{eq:Edefs}. To get the full answer, we also need to include the contribution from the other partition, which is obtained by interchanging 2 and 4 in the previous expression.

Hence, the final answer is obtained as follows
\begin{eqn}
T(-+++) = \text{\eqref{eq:nmhvw+k1k2} + \eqref{eq:nmhvw-k1k2} + \eqref{eq:nmhvk3k4a} + \eqref{eq:nmhvk12} + $(2\leftrightarrow 4)$} .
\end{eqn}

%%%%%%%%%%%%%%%%%%%%%%%%%%
\subsection{MHV case ($T(- + - +)$)  Computation}
\label{sec:mhv1}

\begin{figure}[h]
    \centering
    \includegraphics[width=0.5\linewidth]{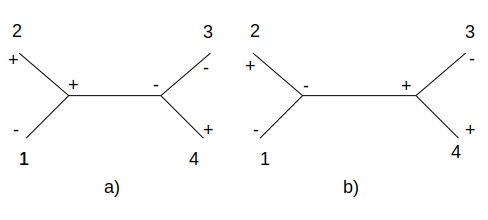}
    \caption{MHV case}
    \label{fig:MHV}
\end{figure}

The deformations for the MHV case ($-+-+$) differ from those of the All-plus and Single-minus cases. As explained in \cite{Raju:2012zs}, the $\bar\lambda$ of the negative helicity gluon is deformed in the following manner,

For $i=1\, \&\,  3$,
\begin{align}
     \lambda_{i}(w) = \lambda_{i},\hskip0.5cm \bar{\lambda}_{i} (w) = \bar{\lambda}_{i} + \beta_{i} w \lambda_{i} .\label{eq:mhvdef1}
\end{align}
The positive helicity gluons follow a pattern similar to the all-plus case,

For $i=2\, \& \,4$,
\begin{align}
    \lambda_{i}(w) = \lambda_{i} + \beta_{i} w \bar{\lambda}_{i}; \, \bar{\lambda}_{i} (w) = \bar{\lambda}_{i} . \label{eq:nmhvdef2}
\end{align}
The procedure for determining $\beta_i$ is exactly the same as the all-plus case, and we obtain,
\begin{align}
    \frac{\beta_2}{\beta_1} = -\frac{\langle \bar{\lambda}_4 \lambda_1 \rangle \, \langle  \lambda_3 \lambda_1 \rangle}{\langle \bar{\lambda}_4 \bar{\lambda}_2 \rangle \, \langle  \lambda_3 \bar{\lambda}_2 \rangle},\, \frac{\beta_3}{\beta_1} = -\frac{\langle \bar{\lambda}_2 \lambda_1 \rangle \, \langle  \bar{\lambda}_4 \lambda_1 \rangle}{\langle \bar{\lambda}_2 \lambda_3 \rangle \, \langle  \bar{\lambda}_4 \lambda_3 \rangle},\, \frac{\beta_4}{\beta_1} = -\frac{\langle \bar{\lambda}_2 \lambda_1 \rangle \, \langle  \lambda_3 \lambda_1 \rangle}{\langle \bar{\lambda}_2 \bar{\lambda}_4 \rangle \, \langle  \lambda_3 \bar{\lambda}_4 \rangle}. \label{eq:betamhv}
\end{align}

As discussed in previous cases, the (12)(34) partition contribution to the four-point correlator integrand is given as: 
\begin{align}
  \hskip-1.cm  I_s(0) &=\sum_{\pm} \frac{-1}{16 \pi |\bm{k_1}||\bm{k_2}| |\bm{k_3}||\bm{k_4}|}  \frac{(|\bm{k_2}| +ip - |\bm{k_1}|)(ip + |\bm{k_1}| -|\bm{k_2}|)(|\bm{k_4}| -ip - |\bm{k_3}|)(-ip + |\bm{k_3}| -|\bm{k_4}|)}{(|\bm{k_1}|+|\bm{k_2}| + ip)((|\bm{k_3}|+|\bm{k_4}| - ip))} \nonumber \\
    &\frac{1}{p^2 + (\bm{k_1} + \bm{k_2})^2}  \frac{w^{\mp}}{w^{\pm} - w^{\mp}}  \left[   \frac{\langle \bar{\lambda}_2 \bar{\lambda}_{int}  \rangle^3}{\langle \bar{\lambda}_{int}  \bar{\lambda}_1 (w) \rangle \langle \bar{\lambda}_1 (w) \bar{\lambda}_2 \rangle} \frac{\langle \lambda_{int} \lambda_3  \rangle^3}{\langle \lambda_3  \lambda_4 (w) \rangle \langle \lambda_4 (w) \lambda_{int} \rangle}   \right. \nonumber \\
    & \left.+ \frac{\langle \lambda_{int} \lambda_1   \rangle^3}{\langle \lambda_1  \lambda_2 (w) \rangle \langle \lambda_2 (w) \lambda_{int} \rangle} \frac{\langle \bar{\lambda}_4 \bar{\lambda}_{int}  \rangle^3}{\langle \bar{\lambda}_{int}  \bar{\lambda}_3 (w) \rangle \langle \bar{\lambda}_3 (w) \bar{\lambda}_4 \rangle}  \right] .\label{eq:MHVbasic}
\end{align}

Here too, we close the contour in the upper p-plane to obtain 3 poles: $p=i(|\bm{k_1}|+|\bm{k_2}|),\,p=i(|\bm{k_3}|+|\bm{k_4}|),\,p= i|\bm{k_1} + \bm{k_2}|$.

\hskip-0.8cm\textbf{(A) For $p=i(|\bm{k_1}|+|\bm{k_2}|)$ pole}

The analysis for this case is exactly the same as the one done for the $p=i(|\bm{k_1}|+|\bm{k_2}|)$ pole of the $T(-+++)$ case. As seen from eq. (\ref{eq:nmhv12w+w-}), here too, we have two solutions for $w$: a) $\langle \lambda_{1} \lambda_{2}(w^+) \rangle=0$ and b) $\langle \bar{\lambda}_{1} (w^-) \bar{\lambda}_{2}\rangle=0$.

\hskip-0.5cm\textbf{i) When $w=w^+$}

The residue is,
\begin{align}
   \hskip-1cm &2\pi i  \underset{\substack{p=i(|\bm{k_1}|+|\bm{k_2}|)\\ w=w^+} }{\operatorname{Res}}\,I_s^{(0)} \nonumber \\
&=  \frac{1}{4  |\bm{k_3}||\bm{k_4}| E}
(|\bm{k_4}| +|\bm{k_1}|+|\bm{k_2}| - |\bm{k_3}|)(|\bm{k_1}|+|\bm{k_2}| + |\bm{k_3}| -|\bm{k_4}|) \nonumber \\
&\hskip0.5cm \times \frac{1}{\langle\lambda_1\lambda_2\rangle} \frac{\langle\lambda_3\lambda_1\rangle^{3}}
     {\langle\lambda_3\lambda_4(w^{+})\rangle\,
      \langle\lambda_4(w^{+})\lambda_2(w^{+})\rangle}  . \label{eq:mhvw+} 
\end{align}

The value of $\lambda_4 (w^+)$ can be obtained by expanding it in terms of $\beta_4 w^+$, which can be written as $\frac{\beta_4}{\beta_2}\,\beta_2 w^+$. We can then substitute values of $\frac{\beta_4}{\beta_2}$ from eq. \eqref{eq:betamhv} and $\beta_2 w^+$ from eq. \eqref{eq:b2w+}. The value of $\lambda_2 (w^+)$ can be obtained by expanding it in terms of $\beta_2 w^+$  and using its expression directly from eq. \eqref{eq:b2w+}.

\hskip-0.5cm\textbf{i) When $w=w^-$}

The residue is,
\begin{align}
   \hskip-1cm &2\pi i  \underset{\substack{p=i(|\bm{k_1}|+|\bm{k_2}|)\\ w=w^-} }{\operatorname{Res}}\,I_s^{(0)} \nonumber \\
 &=   \frac{1}{4  |\bm{k_3}||\bm{k_4}| E}
(|\bm{k_4}| +|\bm{k_1}|+|\bm{k_2}| - |\bm{k_3}|)(|\bm{k_1}|+|\bm{k_2}| + |\bm{k_3}| -|\bm{k_4}|) \nonumber \\
&\hskip0.5cm \times \frac{1}{ \langle\bar{\lambda}_1\bar{\lambda}_2\rangle} \frac{\langle\bar{\lambda}_4\bar{\lambda}_2\rangle^{3}}
     {\langle\bar{\lambda}_3(w^{-})\bar{\lambda}_4\rangle\,
      \langle\bar{\lambda}_3(w^{-})\bar{\lambda}_1(w^{-})\rangle}  .\label{eq:mhvw-}    
\end{align}

The value of $\bar\lambda_3 (w^-)$ can be obtained by expanding it in terms of $\beta_3 w^-$, which can be written as $\frac{\beta_3}{\beta_1}\,\beta_1 w^-$. We can then substitute values of $\frac{\beta_3}{\beta_1}$ from eq. \eqref{eq:betamhv} and $\beta_1 w^-$ from eq. \eqref{eq:b1w-}. The value of $\bar\lambda_1 (w^-)$ can be obtained by expanding it in terms of $\beta_1 w^-$  and using its expression directly from eq. \eqref{eq:b1w-}.

Unlike the All-plus and Single-minus cases, we do get a total energy ($E$)
pole in eqs. \eqref{eq:mhvw+} and \eqref{eq:mhvw-}.

\hskip-0.8cm\textbf{(B) For $p=i(|\bm{k_3}|+|\bm{k_4}|)$ pole}

Similar to the all-plus case, we can interchange $(1,2)\leftrightarrow(3,4)$ in expressions obtained for $p=i(|\bm{k_1}|+|\bm{k_2}|)$ pole above, to obtain the residues for $p=i(|\bm{k_3}|+|\bm{k_4}|)$ pole. This case will, therefore, also contain two separate expressions- one for $w=w^+$ case, and one for $w=w^-$ case. Since we are just interchanging $(1,2)\leftrightarrow(3,4)$ in eqs. \eqref{eq:mhvw+} and \eqref{eq:mhvw-}, we will also get a total energy pole for this residue too, as $E$ will remain invariant under this interchange.

\hskip-0.8cm\textbf{(C) For $p=i(|\bm{k_1}+\bm{k_2}|)$ pole}

Similar to the All-plus and Single-minus cases, the only non-vanishing contribution we get is when $w=0$. The corresponding residue is given as,
\begin{align}
    &\hskip-1cm 2\pi i  \underset{\substack{p=i|\bm{k_1}+\bm{k_2}|} }{\operatorname{Res}}\,I_s^{(0)} \nonumber \\
    &=  \frac{-1}{16 \pi |\bm{k_1}||\bm{k_2}| |\bm{k_3}||\bm{k_4}|}  \frac{(|\bm{k_2}| +ip - |\bm{k_1}|)(ip + |\bm{k_1}| -|\bm{k_2}|)(|\bm{k_4}| -ip - |\bm{k_3}|)(-ip + |\bm{k_3}| -|\bm{k_4}|)}{|\bm{k_1}+\bm{k_2}|(|\bm{k_1}|+|\bm{k_2}| + ip)((|\bm{k_3}|+|\bm{k_4}| - ip))} \nonumber \\
    &\hskip0.5cm \times \Bigg[\frac{\big(E^{12}_{s}\,\langle\lambda_3\bar{\lambda}_2\rangle
      +\langle\lambda_3\lambda_1\rangle\,
       \langle\bar{\lambda}_2\bar{\lambda}_1\rangle\big)^{3}}
     {\langle\bar{\lambda}_1\bar{\lambda}_2\rangle\,
      \langle\lambda_3\lambda_4\rangle\,
      \Big(E^{12}_{s}\,\langle\lambda_4\bar{\lambda}_1\rangle
      +\langle\lambda_4\lambda_2\rangle\,
       \langle\bar{\lambda}_1\bar{\lambda}_2\rangle\Big)} \nonumber \\
      &\hskip1cm + \frac{\big(E^{12}_{s}\,\langle\lambda_1\bar{\lambda}_4\rangle
      +\langle\lambda_1\lambda_2\rangle\,
       \langle\bar{\lambda}_4\bar{\lambda}_2\rangle\big)^{3}}
     {\langle\lambda_1\lambda_2\rangle\,
      \langle\bar{\lambda}_3\bar{\lambda}_4\rangle\,
      \Big(E^{12}_{s}\,\langle\lambda_2\bar{\lambda}_3\rangle
      +\langle\lambda_2\lambda_1\rangle\,
       \langle\bar{\lambda}_3\bar{\lambda}_1\rangle\Big)} \Bigg] .\label{eq:mhvk12}
\end{align}    

$E^{12}_{s}$ is defined in eq. \eqref{eq:Edefs}. Like in the All-plus and Single-minus cases, to get the full answer, we also need to include the contribution from the other partition, which is obtained by interchanging 2 and 4 in the previous expression.

Hence, the final answer is obtained as follows,
\begin{eqn}
T(-+-+) = \text{\eqref{eq:mhvw+} + \eqref{eq:mhvw-} +  \Big\{\eqref{eq:mhvw+} + \eqref{eq:mhvw-}\Big\}$\Big|_{(1,2)\leftrightarrow (3,4)}$ + \eqref{eq:mhvk12} + $(2\leftrightarrow 4)$} .
\end{eqn}

Note that, \cite{Raju:2012zs} computed $T(+-+-)$, whereas we compute $T(-+-+)$. Generally, $T(+-+-)$ and $T(-+-+)$ are not equivalent. But in this case, $T(+-+-)$ calculated by \cite{Raju:2012zs} matches our result because of a different definition of spinors (eq. (2.5) of \cite{Raju:2012zs}) used by \cite{Raju:2012zs}. We find agreement up to a minor typographical error \footnote{There is a missing factor of $ \frac{1}{ |\bm{k_1}||\bm{k_2}| |\bm{k_3}||\bm{k_4}|}$ in eq.(5.13) of \cite{Raju:2012zs}.}.

%%%%%%%%%%%%%%%%%%%%%%%%%%
\section{Numerical Checks}
\label{sec:check}
%%%%%%%%%%%%%%%%%%%%%%%%%%

We calculate the values of the four-point correlator using the expressions for all three cases: (i) All-plus $T(++++)$, (ii) Single-minus $T(-+++)$, and (iii) MHV $T(-+-+)$, obtained following all-line recursion relations, as described in section {\ref{sec:bcfw}}. We perform this computation for two fixed sets of momentum values such that momentum conservation is satisfied.
We also calculate the values of the correlators obtained via Witten diagram computation. For this, we use the expressions for All-plus $T(++++)$, Single-minus $T(-+++)$, and MHV $T(-+-+)$ given in eqs. \eqref{eq:almallp}, \eqref{eq:almnmhv}, and \eqref{eq:almmhv} of \cite{ALM:2021}, respectively. We find that the values of the correlators obtained via Witten diagram computation match completely with the values obtained via the all-line recursion relations. Table \ref{table:1} demonstrates this. 

\begin{table}[h!]
\centering
\begin{tabular}{| c| c | c | c |} 
 \hline
 The Correlator & Momenta Set & Witten Diagram REF & All-line Recursion Relations  \\ 
 \hline\hline
 $T(++++)$ & A & $0.0110056 + i0.0186617$  & $0.0110056 + i0.0186617$  \\ 
           & B &  $-0.0259340 -i 0.0294247$  &  $-0.0259340 -i 0.0294247$   \\
 \hline          
 $T(-+++)$ & A & $-0.0604496- i 0.0453051$ & $-0.0604496- i 0.0453051$  \\
            & B &  $-0.2006446-i 0.0356869$  & $-0.2006446-i 0.0356869$   \\
 \hline 
 $T(-+-+)$ & A & $0.6680843 - i 0.1951043$ &  $0.6680843 - i 0.1951043$ \\
            & B &  $0.2603628-i 0.0633525$  &   $0.2603628-i 0.0633525$  \\
 \hline
\end{tabular}
\caption{The first column shows all the 3 four-point correlators considered in this study: (i) All-plus  $T(++++)$, (ii) Single-minus $T(-+++)$, and (iii) MHV $T(-+-+)$. The second column shows the values of each of the correlators using the expressions derived from Witten diagram computation. The third column shows the values of each of the correlators using the expressions derived from the all-line recursion relations. Here, set A refers to $\bm{k_1}=(1,0.2,0)$, $\bm{k_2}=(-0.3,1.1,0)$, $\bm{k_3}=(-0.8,-0.4,0)$, and $\bm{k_4}=(0.1,-0.9,0)$; set B refers to $\bm{k_1}=(1,0.2,0.5)$, $\bm{k_2}=(-0.3,1.1,-0.7)$, $\bm{k_3}=(-0.8,-0.4,0.9)$, and $\bm{k_4}=(0.1,-0.9,-0.7)$. The values obtained from both formalisms for corresponding correlators completely match.}
\label{table:1}
\end{table}
Note that the three-point vertex expression used by \cite{ALM:2021} in their eq. (86), differs by a factor of $-i$ from the three-point vertex expression used by \cite{Raju:2012zs} as well as us. \cite{ALM:2021} explicitly mention that they have dropped the factor of $i$. Hence, for the four-point function, due to this discrepancy of $(-i)^2$, the values of the correlator from \cite{ALM:2021} lack a minus sign. We have taken into account this minus sign while calculating the correlator values via the Witten diagram computations (i.e., Witten diagram four-point correlator values in the table = -1$\times$ (correlator values obtained from \cite{ALM:2021})), and hence the values match exactly with the values of the correlators derived using our results.

\section{Conclusions and Future Scope}
\label{sec:conclusions}

In this work, we have used the all-line recursion relations of
\cite{Raju:2012zr,Raju:2012zs} to compute the four-point Yang-Mills
color-ordered current correlator in $\mathrm{AdS}_{4}/\mathrm{CFT}_{3}$ for (i) All-plus  $T(++++)$, (ii) Single-minus $T(-+++)$, and (iii) MHV $T(-+-+)$ helicity assignments. In each case the answer is a finite sum of residues at the three poles, summed over the two color-ordered partitions. We give the explicit derivation of all the residues for each of the three correlators. We also find that for the All-plus and Single-minus cases we do not get a total energy ($E$) pole, whereas for the MHV case we do get a total energy pole. 

The central claim of this paper, in addition to the formulae being derived using all-line recursion relations for these specific helicity assignments, is also that they have been tested against the Witten diagram computation results for the same correlators. We have checked that, for all the cases: (i) All-plus $T(++++)$, (ii) Single-minus $T(-+++)$, and (iii) MHV $T(-+-+)$, our results derived using all-line recursion relations numerically match the Witten diagram computation results. We show the equivalence for two sets of external momenta as examples in Table \ref{table:1}. The only systematic discrepancy with the results obtained using \cite{ALM:2021} correlator values was an overall sign, traced to the $-i$ relative normalisation of the cubic vertex.

Several directions remain to be explored. One is the application of the all-line recursion relation for computing higher point functions, such as five-point functions, six-point functions, and n-point functions. A five-point correlator can be decomposed into products of three-point and four-point transition amplitudes. Likewise, a six-point correlator can be decomposed into products containing either one three-point and one five-point transition amplitude, or two four-point amplitudes. Specifically, the all-plus five-point correlator $T_5(+++++)$, will be obtained from calculating $T_3(+,+,+_{int})\,T_4(-_{int},+,+,+) + T_3(+,+,-_{int})\,T_4(+_{int},+,+,+)$. Thus, the all-plus five-point correlator will not require any MHV four-point transition amplitude. Similarly, the all-plus six-point correlator $T_6(+++++)$, will require calculation of (i) $T_3(+,+,+_{int})\,T_5(-_{int},+,+,+,+) + T_3(+,+,-_{int})\,T_5(+_{int},+,+,+,+)$, and (ii) $T_4(+,+,+,+_{int})\,T_4(-_{int},+,+,+) + T_4(+,+,+,-_{int})\,T_4(+_{int},+,+,+)$. 
One important thing to consider here is that $T_4(-_{int},+,+,+)$, $T_4(+_{int},+,+,+)$ and other sub-amplitudes mentioned above are transition amplitudes and not vacuum correlators. This means that, while computing $T_4(+_{int},+,+,+)$ recursively, we will encounter a cubic vertex that will have only one external leg and two normalizable modes attached to it, unlike the correlators discussed in this work, where we had two external legs at each cubic vertex. Apart from the propagator, the rest of the poles are provided by the leading factor of the cubic vertex. At the cubic vertex with one external leg, the leading factor will be $\propto \int_{0}^{\infty} z^{3/2} K_{1/2}(|\bm{k_i}|z) J_{1/2}(pz) J_{1/2}(p'z) \,dz$, where $K_{1/2}(|\bm{k_i}|z)$ denotes the bulk-to-boundary propagator corresponding to the $i^{th}$ external leg, $J_{1/2}(pz)$ and $J_{1/2}(p'z)$ are bulk-to-bulk propagators corresponding to the two normalizable modes. Though such structures are studied in the context of Witten diagrams \cite{Albayrak:2023jzl}, we will explore them further in the context of recursion relations in our future paper \cite{Chowdhury-Pathak-et.al}.

The AdS BCFW recursion relations trivially work for the wavefunction in dS, as the dS wavefunction is related to the AdS correlators by a simple analytic continuation. In principle, one can also set up a recursion relation for directly obtaining the correlator in dS. The dS correlators of the gauge field are only well defined when the choice of the gauge fixing at the boundary is specified. We revisit this in more detail in a future paper \cite{Chowdhury-Pathak-et.al}. We will also work on the natural extension of this work to the graviton correlators in future work.

%%%%%%%%%%%%%%%%%%%%%%%%%%
\subsection*{Acknowledgements}
We are grateful to Chandramouli Chowdhury for their immensely useful discussions and insights into the manuscript. 
%%%%%%%%%%%%%%%%%%%%%%%%%%
%%%%%%%%%%%%%%%%%%%%%%%%%%
%%%%%%%%%%%%%%%%%%%%%%%%%%
\begin{appendix}
%%%%%%%%%%%%%%%%%%%%%%%%%%
\section{Spinor Helicity Formalism in AdS}
\label{sec:spinform}
%%%%%%%%%%%%%%%%%%%%%%%%%%
In this section, we give a brief review of the notations and conventions as well as the spinor helicity formalism used in this paper.

The antisymmetric matrices used in this work are defined as,
\begin{align}
  e_{\alpha\beta} = e_{\dot{\alpha}\dot{\beta}} = e_{\alpha\dot{\beta}}
=
\begin{pmatrix}
0 & 1 \\
-1 & 0
\end{pmatrix}
=
-e_{\beta\alpha}
=
-e_{\dot{\beta}\dot{\alpha}} = -e_{\dot{\beta}\alpha}=-e^{\alpha\beta} = -e^{\dot{\alpha}\dot{\beta}} = -e^{\alpha\dot{\beta}}. 
\end{align}

Different relations between momenta and spinors used in this work are,
\begin{align}
     & \sum_{i} \bm{k_i} = 0 , \quad
    \sum_{i} \lambda_{i}^{\alpha} \bar{\lambda}_{i}^{\dot{\alpha}} = -E\,e^{\alpha \dot{\alpha}}, \quad
      \langle \bar{\lambda}_{i}\lambda_{i}\rangle = 2|\bm{k_i}|, \quad \sum_{i} \langle \bar{\lambda}_{i}\lambda_{i}\rangle = \sum_{i} 2|\bm{k_i}| = 2E , \quad \sum_{i} \langle \lambda_{i} \bar{\lambda}_{i}\rangle = -2E .
\end{align}

The spinor representation of polarization vectors is given as,
\begin{align}
    \epsilon^{+ \alpha \dot{\alpha}}_i &= 2 \frac{\bar{\lambda}_i^{\alpha}  \bar{\lambda}_i^{\dot{\alpha}}}{\langle \bar{\lambda}_i \lambda_i \rangle} = \frac{\bar{\lambda}_i^{\alpha}  \bar{\lambda}_i^{\dot{\alpha}}}{|\bm{k_i}|} ,\\
     \epsilon^{- \alpha \dot{\alpha}}_i &= -2 \frac{\lambda_i^{\alpha}  \lambda_i^{\dot{\alpha}}}{\langle \bar{\lambda}_i \lambda_i \rangle} = -\frac{\lambda_i^{\alpha}  \lambda_i^{\dot{\alpha}}}{|\bm{k_i}|} .
\end{align}

The convention used in this paper for raising and lowering of indices can be understood from the following expression of the mixed spinor product,
\begin{align}
    \langle \bar{\lambda}_1 \lambda_2\rangle = \bar{\lambda}_{1,\alpha} \lambda_2^{\alpha} = e_{\alpha \dot{\alpha}} \bar{\lambda}_{1}^{\dot{\alpha}} \lambda_2^{\alpha} = e_{\alpha \dot{\alpha}} e^{\dot{\alpha} \dot{\beta}} \bar{\lambda}_{1,\dot{\beta}} \lambda_2^{\alpha} = \delta^{\dot{\beta}}_{\alpha} \bar{\lambda}_{1,\dot{\beta}} \lambda_2^{\alpha}  =  \bar{\lambda}_{1,\dot{\beta}} \lambda_2^{\dot{\beta}} .
\end{align}

For the (12)(34) partition in the four-point case, the relation between the mixed product of the intermediate spinors and $p$ is given as,
\begin{align}
    \langle \bar{\lambda}_{int} \lambda_{int}\rangle = 2|\bm{k_{int}}| = 2ip .
\end{align}

For $a,b=1,2,3,4$, we also have the following relations,
\begin{align}
    &(|\bm k_a|+|\bm k_b|)^2-(\bm k_a+\bm k_b)^2=\langle \lambda_a \lambda_b \rangle \langle \bar\lambda_a \bar\lambda_b \rangle , \label{eq:llblbl}\\
    &|\bm{k_a}| + |\bm{k_b}| \pm ip = E^{ab}_{\pm p}. \label{eq:Eabpdef}
\end{align}

We also have the following relations between spinors,
\begin{align}
    &\lambda_{1}^{\alpha} \bar{\lambda}_{1}^{\dot{\alpha}} + \lambda_{2}^{\alpha} \bar{\lambda}_{2}^{\dot{\alpha}} + \lambda_{int}^{\alpha} \bar{\lambda}_{int}^{\dot{\alpha}} = -E^{12}_{p} e^{\alpha \dot{\alpha}}, \\
     &\lambda_{3}^{\alpha} \bar{\lambda}_{3}^{\dot{\alpha}} + \lambda_{4}^{\alpha} \bar{\lambda}_{4}^{\dot{\alpha}} - \lambda_{int}^{\alpha} \bar{\lambda}_{int}^{\dot{\alpha}} = -E^{34}_{-p} e^{\alpha \dot{\alpha}} ,\\   
    &\lambda_{1}^{\alpha} \bar{\lambda}_{1}^{\dot{\alpha}} + \lambda_{2}^{\alpha} \bar{\lambda}_{2}^{\dot{\alpha}} + \lambda_{3}^{\alpha} \bar{\lambda}_{3}^{\dot{\alpha}} + \lambda_{4}^{\alpha} \bar{\lambda}_{4}^{\dot{\alpha}} = -E e^{\alpha \dot{\alpha}} \label{eq:allspinorsum}.
\end{align}

To express the intermediate spinors in terms of external spinors, we perform contractions to get,
\begin{align}
&\langle \lambda_m \lambda_{int}\rangle \langle \bar\lambda_n \bar\lambda_{int}\rangle =  -E^{12}_{p} \langle \lambda_m \bar\lambda_n \rangle -  \langle \lambda_m \lambda_1\rangle \langle \bar\lambda_n \bar\lambda_1\rangle  - \langle \lambda_m \lambda_2\rangle \langle \bar\lambda_n \bar\lambda_2\rangle , \\
&\langle \lambda_m \lambda_{int}\rangle \langle \bar\lambda_n \bar\lambda_{int}\rangle =  E^{34}_{-p} \langle \lambda_m \bar\lambda_n \rangle +  \langle \lambda_m \lambda_3\rangle \langle \bar\lambda_n \bar\lambda_3\rangle  + \langle \lambda_m \lambda_4\rangle \langle \bar\lambda_n \bar\lambda_4\rangle .
    \label{eq:intspinor}
\end{align}

We consider the three-momentum of particle $i$ as $\bm{k_i} = (k_{i,1},k_{i,2},k_{i,3})$. The magnitude of the momentum is given $|\bm{k_i}|  = \sqrt{k_{i,1}^2 + k_{i,2}^2 + k_{i,3}^2}$. The spinors are expressed in terms of momentum components as follows,
\begin{align}
    \lambda_i = \left( \sqrt{|\bm{k_i}| + k_{i,3}}, \, \frac{k_{i,1} - ik_{i,2}}{\sqrt{|\bm{k_i}| + k_{i,3}}}  \right);\,\bar\lambda_i = \left(\frac{-k_{i,1} - ik_{i,2}}{\sqrt{|\bm{k_i}| + k_{i,3}}}, \sqrt{|\bm{k_i}| + k_{i,3}}\right) .
    \label{eq:lamdef}
\end{align}

Some definitions:
\begin{align}
    &E^{12}_{s}= |\bm{k_1}|+|\bm{k_2}| -|\bm{k_1} + \bm{k_2}|,\hskip0.5cm E^{34}_{-s}= |\bm{k_3}|+|\bm{k_4}| +|\bm{k_1} + \bm{k_2}| =  |\bm{k_3}|+|\bm{k_4}| +|\bm{k_3} + \bm{k_4}| , \\
    &E^{34}_{-12}= |\bm{k_3}|+|\bm{k_4}|+ |\bm{k_1}|+|\bm{k_2}| = E,\hskip0.5cm  E^{12}_{34} = |\bm{k_1}|+|\bm{k_2}| - |\bm{k_3}|-|\bm{k_4}| . \label{eq:Edefs}
\end{align}

\section{Three-point functions}
\label{sec:3point}

We use the following expressions for the three-point functions,
\begin{align}
    T_{3}(++-) = \frac{R^{YM}(|\bm{k_1}|,|\bm{k_2}|,p)}{2\sqrt{2}|\bm{k_1}||\bm{k_2}|p} \left(|\bm{k_2}| + ip -|\bm{k_1}|\right) \left(ip + |\bm{k_1}| - |\bm{k_2}| \right) \left(|\bm{k_1}|+|\bm{k_2}|-ip\right) \frac{\langle \bar\lambda_1 \bar\lambda_2\rangle^3}{\langle \bar\lambda_2 \bar\lambda_3 \rangle \langle \bar\lambda_3 \bar\lambda_1\rangle},
\end{align}

\begin{align}
    T_{3}(+++) = \frac{R^{YM}(|\bm{k_1}|,|\bm{k_2}|,p)}{2\sqrt{2}|\bm{k_1}||\bm{k_2}|p} E^{12}_{p} \langle \bar\lambda_1 \bar\lambda_2\rangle \langle \bar\lambda_2 \bar\lambda_3\rangle \langle \bar\lambda_3 \bar\lambda_1\rangle ,
\end{align}

\begin{align}
    T_{3}(--+) = \frac{R^{YM}(|\bm{k_1}|,|\bm{k_2}|,p)}{2\sqrt{2}|\bm{k_1}||\bm{k_2}|p} \left(|\bm{k_2}| + ip -|\bm{k_1}|\right) \left(ip + |\bm{k_1}| - |\bm{k_2}| \right) \left(|\bm{k_1}|+|\bm{k_2}|-ip\right) \frac{\langle \lambda_1 \lambda_2\rangle^3}{\langle \lambda_2 \lambda_3 \rangle \langle \lambda_3 \lambda_1\rangle} ,
\end{align}

\begin{align}
    T_{3}(---) = \frac{R^{YM}(|\bm{k_1}|,|\bm{k_2}|,p)}{2\sqrt{2}|\bm{k_1}||\bm{k_2}|p} E^{12}_{p} \langle \lambda_1 \lambda_2\rangle \langle \lambda_2 \lambda_3\rangle \langle \lambda_3 \lambda_1\rangle .
\end{align}

NOTE: We use the expression of $R^{YM}(|\bm{k_1}|, |\bm{k_2}|, p)$ as derived  in eq. (3.7) of \cite{Raju:2012zs}-
\begin{align}
\label{fprefact}
R^{YM}(|\bm{k_1}|, |\bm{k_2}|, p) &= { 2 \sqrt{|\bm{k_1}| |\bm{k_2}|}  \over \pi} \int_0^{\infty} z^{3 \over 2} K_{1/2}(|\bm{k_1}| z) K_{1/2}( |\bm{k_2}| z) J_{1/2} (p z) d z \nonumber \\  
&= \frac{\sqrt{\frac{2 p }{\pi }}}{|\bm{k_1}|^2+2 |\bm{k_2}| |\bm{k_1}|+|\bm{k_2}|^2+p^2}.
\end{align}

\section{Derivations of the (12)(34) partition residues}
\label{app:schan}

While evaluating the residues for any of the three cases $T(++++)$, $T(-+++)$, or $T(-+-+)$, we see that $w$ satisfies a quadratic equation. The two solutions of this quadratic equation ($w^+$ and $w^-$) can be expressed in terms of spinor products. Depending upon the helicities, and consequently the deformations, we deal with the elimination of $w^{\pm}$ differently, which we explore in this section. 

\subsection{For $p=i(|\textbf{k}_1|+|\textbf{k}_2|)$ pole}
\label{sec:k1k2}
In the case of the pole $p=i(|\bm{k_1}|+|\bm{k_2}|)$, for all three four-point functions considered in this paper,  we have $|\bm{k_1}|+|\bm{k_2}| + ip = E^{12}_{p} = 0$. This gives,
\begin{align}
& \lambda_{1}^{'\alpha} \bar{\lambda}_{1}^{\dot{'\alpha}} + \lambda_{2}^{'\alpha} \bar{\lambda}_{2}^{'\dot{\alpha}} =-\lambda_{int}^{\alpha} \bar{\lambda}_{int}^{\dot{\alpha}}, \label{eq:k1k2int} \\%
   \implies&  \left(\lambda_{1}^{'\alpha} \bar{\lambda}_{1}^{'\dot{\alpha}} + \lambda_{2}^{'\alpha} \bar{\lambda}_{2}^{'\dot{\alpha}}\right) e_{\alpha \beta}e_{\dot{\alpha} \dot{\beta}} \left(\lambda_{1}^{'\beta} \bar{\lambda}_{1}^{'\dot{\beta}} + \lambda_{2}^{'\beta}  \bar{\lambda}_{2}^{\dot{'\beta}}\right) = 0 \hskip1cm[\text{On self contracting, RHS vanishes}], \nonumber \\
   \implies& \langle \lambda'_{1} \lambda'_{2} \rangle \langle \bar{\lambda}'_{1} \bar{\lambda}'_{2}\rangle = 0. \label{eq:k1k2A}
\end{align}

For $T(++++)$, $\lambda'_{1} = \lambda_{1}(w),\,\lambda'_{2} = \lambda_{2}(w),\,\bar\lambda'_{1} = \bar\lambda_{1},\,\bar\lambda'_{2} = \bar\lambda_{2}$. Therefore, from eq. (\ref{eq:k1k2A}) we get, $\langle \lambda_{1}(w) \lambda_{2} (w) \rangle=0$. This gives,
\begin{align}
    &\langle \lambda_{1}(w) \lambda_{2} (w) \rangle =  \langle \lambda_{1} + \beta_{1} w \bar{\lambda}_{1} | \lambda_{2} + \beta_{2} w \bar{\lambda}_{2} \rangle =0 , \nonumber \\
    &\implies  w^2 = -\frac{\langle \lambda_{1} \lambda_{2} \rangle}{\beta_{1} \beta_{2} \langle \bar{\lambda}_{1} \bar{\lambda}_{2} \rangle}- \frac{(\beta_1 \langle \bar{\lambda}_{1} \lambda_{2} \rangle  +\beta_2 \langle  \lambda_{1} \bar{\lambda}_{2} \rangle )}{\beta_{1} \beta_{2} \langle \bar{\lambda}_{1} \bar{\lambda}_{2} \rangle} w .\label{eq:wquadallpk1k2}   
\end{align}
Since, we have $\langle \lambda_{1}(w) \lambda_{2} (w) \rangle=0$, this also implies, $\lambda_2(w)=t(w)\lambda_1(w)$. Substituting $\lambda_2(w)$ in eq. (\ref{eq:k1k2int}), we get,
\begin{align}
&\lambda_{1}^{\alpha}(w) \left(\bar{\lambda}_{1}^{\dot{\alpha}} + t(w) \bar{\lambda}_{2}^{\dot{\alpha}} \right) =-\lambda_{int}^{\alpha} \bar{\lambda}_{int}^{\dot{\alpha}}, \nonumber \\
\implies & \lambda_{\rm int} = \lambda_1 (w); \bar{\lambda}_{\rm int} = -\left(\bar{\lambda}_1+t(w)\bar{\lambda}_2\right). \label{eq:lintrelk1k2allp}
\end{align}
Additionally, using the definition of deformations for the all-plus case in eq. (\ref{eq:allpdeform}), and carrying out contractions, we can express $t(w)$ and $1/t(w)$ as,
\begin{align}
 &t(w) = \frac{\langle \bar{\lambda}_1 \lambda_2 \rangle + \beta_2 w \langle \bar{\lambda}_1 \bar{\lambda}_2 \rangle}{2|\bm{k_1}|} \hskip0.5cm[\text{from }\langle \bar{\lambda}_1 \lambda_2(w) \rangle=\langle \bar{\lambda}_1 t(w) \,\lambda_1(w) \rangle] , \\
  &\frac{1}{t(w)} = \frac{ \langle \bar{\lambda}_2 \lambda_1 \rangle + \beta_1 w \langle \bar{\lambda}_2 \bar{\lambda}_1\rangle}{ 2|\bm{k_2}|} \hskip0.5cm[\text{from }\langle \bar{\lambda}_2 \lambda_2(w) \rangle  = \langle \bar{\lambda}_2 \,t(w)\lambda_1(w) \rangle] .
\end{align}

The expression for the residue for the $p=i(|\bm{k_1}|+|\bm{k_2}|)$ starting from eq. (\ref{eq:allpschan}), after some algebra, is given by,
\begin{align}
    \hskip-1cm &2\pi i\underset{p=i(|\bm{k_1}|+|\bm{k_2}|)}{\operatorname{Res}}\,I_s^{(0)} \nonumber \\
     &=  \frac{1}{8 |\bm{k_1}||\bm{k_2}| |\bm{k_3}||\bm{k_4}|}  \frac{(2|\bm{k_1}|)(2|\bm{k_2}|)}{(|\bm{k_3}|+|\bm{k_4}| - |\bm{k_1}|-|\bm{k_2}|)}  \frac{\langle \bar{\lambda}_1 \bar{\lambda}_2 \rangle^2 \langle \bar{\lambda}_3 \bar{\lambda}_4  \rangle }{\langle \lambda_1 \lambda_2 \rangle} \sum_{\pm} \frac{\langle \bar{\lambda}_4 \bar{\lambda}_{int}\rangle \langle \bar{\lambda}_{int} \bar{\lambda}_3 \rangle}{\langle \bar{\lambda}_2  \bar{\lambda}_{int} \rangle \langle \bar{\lambda}_{int} \bar{\lambda}_1 \rangle}  \frac{w^{\mp}}{w^{\pm} - w^{\mp}} . \label{eq:allpk1k2}
\end{align}

Substituting, the expression for $\bar{\lambda}_{\rm int}$ from eq. (\ref{eq:lintrelk1k2allp}) in factor $\frac{\langle \bar{\lambda}_4 \bar{\lambda}_{int}\rangle \langle \bar{\lambda}_{int} \bar{\lambda}_3 \rangle}{\langle \bar{\lambda}_2  \bar{\lambda}_{int} \rangle \langle \bar{\lambda}_{int} \bar{\lambda}_1 \rangle}$ of eq. (\ref{eq:allpk1k2}), we get,
\begin{align}
\frac{ \langle \bar{\lambda}_4 \bar{\lambda}_{\rm int} \rangle \langle \bar{\lambda}_{\rm int} \bar{\lambda}_3 \rangle}{\langle \bar{\lambda}_2 \bar{\lambda}_{\rm int} \rangle \langle \bar{\lambda}_{\rm int} \bar{\lambda}_1 \rangle} &= -\frac{1}{\langle \bar{\lambda}_1 \bar{\lambda}_2 \rangle^2} \left[\langle \bar{\lambda}_1 \bar{\lambda}_4 \rangle \langle \bar{\lambda}_2 \bar{\lambda}_3 \rangle + \langle \bar{\lambda}_2 \bar{\lambda}_4 \rangle \langle \bar{\lambda}_1 \bar{\lambda}_3 \rangle + \frac{
\langle \bar{\lambda}_1 \bar{\lambda}_4 \rangle \langle \bar{\lambda}_1 \bar{\lambda}_3 \rangle}{t(w)} + t(w) \langle \bar{\lambda}_2 \bar{\lambda}_4 \rangle \langle \bar{\lambda}_2 \bar{\lambda}_3 \rangle \right] \nonumber \\
&= -\frac{1}{\langle \bar{\lambda}_1 \bar{\lambda}_2 \rangle^2} \left[ \langle \bar{\lambda}_1 \bar{\lambda}_4 \rangle \langle \bar{\lambda}_2 \bar{\lambda}_3 \rangle + \langle \bar{\lambda}_2 \bar{\lambda}_4 \rangle \langle \bar{\lambda}_1 \bar{\lambda}_3 \rangle - \frac{ \langle \bar{\lambda}_2 \bar{\lambda}_4 \rangle \langle \bar{\lambda}_2 \bar{\lambda}_3 \rangle \langle \lambda_2 \bar{\lambda}_1 \rangle}{2|\bm{k_1}|} - \frac{ \langle \bar{\lambda}_1 \bar{\lambda}_4 \rangle \langle \bar{\lambda}_1 \bar{\lambda}_3 \rangle \langle \lambda_1 \bar{\lambda}_2 \rangle}{2|\bm{k_2}|}\right] \nonumber \\
&\hskip3cm -\frac{w}{\langle \bar{\lambda}_1 \bar{\lambda}_2 \rangle^2}\Bigg[  \frac{ \langle \bar{\lambda}_2 \bar{\lambda}_4 \rangle \langle \bar{\lambda}_2 \bar{\lambda}_3 \rangle \langle  \bar{\lambda}_1 \bar{\lambda}_2  \rangle}{2|\bm{k_1}|} + \frac{ \langle \bar{\lambda}_1 \bar{\lambda}_4 \rangle \langle \bar{\lambda}_1 \bar{\lambda}_3 \rangle \langle \bar{\lambda}_2 \bar{\lambda}_1 \rangle}{2|\bm{k_2}|} \Bigg] .\label{eq:allpintfack1k2}
\end{align}

Thus eq. (\ref{eq:allpintfack1k2}) is linear in $w_\pm$. For any linear function
$f(w)=f_0+f_1w$, the root weight used satisfies, 
\begin{equation}
\sum_{\pm}
f(w_\pm)
\frac{w_\mp}{w_\pm-w_\mp}
=
\frac{
w_-f(w_+)-w_+f(w_-)
}{
w_+-w_-
}
=
-f_0.
\label{eq:fwf0rel}
\end{equation}

Hence, as per eq. (\ref{eq:allpintfack1k2}), 
\begin{align}
    f_0 = -\frac{1}{\langle \bar{\lambda}_1 \bar{\lambda}_2 \rangle^2} \left[ \langle \bar{\lambda}_1 \bar{\lambda}_4 \rangle \langle \bar{\lambda}_2 \bar{\lambda}_3 \rangle + \langle \bar{\lambda}_2 \bar{\lambda}_4 \rangle \langle \bar{\lambda}_1 \bar{\lambda}_3 \rangle - \frac{ \langle \bar{\lambda}_2 \bar{\lambda}_4 \rangle \langle \bar{\lambda}_2 \bar{\lambda}_3 \rangle \langle \lambda_2 \bar{\lambda}_1 \rangle}{2|\bm{k_1}|} - \frac{ \langle \bar{\lambda}_1 \bar{\lambda}_4 \rangle \langle \bar{\lambda}_1 \bar{\lambda}_3 \rangle \langle \lambda_1 \bar{\lambda}_2 \rangle}{2|\bm{k_2}|}\right]. \label{eq:f0allpk1k2}
\end{align}

Using eq. (\ref{eq:f0allpk1k2}), eq. (\ref{eq:allpk1k2}), and applying the Schouten identity, we get the residue contribution of $p=i(|\bm{k_1}|+|\bm{k_2}|)$ pole to $T(++++)$.

For $T(-+++)$ and $T(-+-+)$, $\lambda'_{1} = \lambda_{1},\,\lambda'_{2} = \lambda_{2}(w),\,\bar\lambda'_{1} = \bar\lambda_{1}(w),\,\bar\lambda'_{2} = \bar\lambda_{2}$. Therefore, from eq. (\ref{eq:k1k2A}) we get, $\langle \lambda_{1} \lambda_{2}(w) \rangle \langle \bar{\lambda}_{1} (w)\bar{\lambda}_{2}\rangle = 0$. This gives us two solutions for $w$: 
\begin{align}
    \langle \lambda_{1} \lambda_{2}(w^+) \rangle=0, \hskip0.5cm \langle \bar{\lambda}_{1} (w^-) \bar{\lambda}_{2}\rangle=0. \label{eq:w+w-A}
\end{align}

\hskip-0.8cm\textbf{i) When $w=w^+$:}\\
For $\langle \lambda_{1} \lambda_{2}(w^+) \rangle=0$ case, expanding $\lambda_{2}(w^+)$ in terms of $\beta_2$ and $w^+$ gives,
\begin{align}
\beta_2 w^+ = -\frac{\langle \lambda_1 \lambda_2 \rangle}{\langle \lambda_1 \bar{\lambda}_2 \rangle} . \label{eq:b2w+}
\end{align}

While computing the residue at $w=w^+$, we can write,
\begin{align}
    \langle \bar{\lambda}_1 (w^+)\bar{\lambda}_2 \rangle &=  \langle \bar{\lambda}_1 \bar{\lambda}_2 \rangle +  \langle \lambda_1 \bar{\lambda}_2 \rangle \beta_1 w^- \frac{w^+}{w^-} = \langle \bar{\lambda}_1 \bar{\lambda}_2 \rangle \,\frac{w^- - w^+}{w^-} \hskip0.5cm[\text{Using eq. (\ref{eq:b1w-})}] . \label{eq:1w+2wrel}
\end{align} 
We can combine the propagator and the $\frac{w^{\mp}}{w^{\pm}-w^{\mp}}$ factor using eq. (\ref{eq:1w+2wrel}) as,
\begin{align}
    \frac{1}{p^2 + (\bm{k_1} + \bm{k_2})^2} \frac{w^{-}}{w^{+}-w^{-}} &=\frac{1}{(|\bm{k_1}|+|\bm{k_2}|)^2 - (\bm{k_1} + \bm{k_2})^2} \frac{w^{-}}{w^{-}-w^{+}} , \nonumber \\
    &=\frac{1}{\langle \lambda_1 \lambda_2 \rangle \langle \bar{\lambda}_1 \bar{\lambda}_2 \rangle} \frac{w^{-}}{w^{-}-w^{+}} , \nonumber \\
    &= \frac{1}{\langle \lambda_1 \lambda_2 \rangle \langle \bar{\lambda}_1 (w^+)  \bar{\lambda}_2 \rangle} . \label{eq:w+k12l1l2}
\end{align}

Additionally, $\langle \lambda_{1} \lambda_{2}(w^+) \rangle=0$ implies $\lambda_{2}(w^+) = C_1 \lambda_{1}$. We know,
\begin{align}
    & \lambda_{1}^{\alpha} \bar{\lambda}_{1}^{\dot{\alpha}} (w) + \lambda_{2}^{\alpha} (w) \bar{\lambda}_{2}^{\dot{\alpha}} =-\lambda_{int}^{\alpha} \bar{\lambda}_{int}^{\dot{\alpha}} , \nonumber \\
 \implies   & \lambda_{1}^{\alpha} (\bar{\lambda}_{1}^{\dot{\alpha}} (w) + C_1 \bar{\lambda}_{2}^{\dot{\alpha}}) = -\lambda_{int}^{\alpha} \bar{\lambda}_{int}^{\dot{\alpha}} , \nonumber \\
 \implies & \lambda_{1}=\lambda_{int} \hskip1cm[\text{comparing both sides}] .
\end{align}

For $T(-+++)$, under these conditions, we can see that the nonvanishing contribution to the residue comes from diagram (b) of Fig. \ref{fig:NMHV}.

\hskip-0.8cm\textbf{ii) When $w=w^-$:}\\
For $\langle \bar{\lambda}_1(w^-) \bar{\lambda}_2 \rangle = 0$ case, expanding $\bar\lambda_{1}(w^-)$ in terms of $\beta_1$ and $w^-$ gives,
\begin{align}
 \beta_1 w^- = -\frac{\langle \bar{\lambda}_1 \bar{\lambda}_2 \rangle}{\langle \lambda_1 \bar{\lambda}_2 \rangle} . \label{eq:b1w-}
\end{align}

On the other hand, while computing the residue at $w=w^-$, we can write, 
\begin{align}
    \langle \lambda_1 \lambda_2 (w^-) \rangle =  \langle \lambda_1 \lambda_2 \rangle +  \langle \lambda_1 \bar{\lambda}_2 \rangle \beta_2 w^+ \frac{w^-}{w^+} = \langle \lambda_1 \lambda_2  \rangle \,\frac{w^+ - w^-}{w^+} \hskip0.5cm[\text{Using eq. (\ref{eq:b2w+})}] .\label{eq:1w-2wrel}
\end{align}
Here too, we combine the propagator and the $\frac{w^{\mp}}{w^{\pm}-w^{\mp}}$ factor using eq. (\ref{eq:1w-2wrel}), by following steps similar to those used in obtaining eq. (\ref{eq:w+k12l1l2}). We get,
\begin{align}
    \frac{1}{p^2 + (\bm{k_1} + \bm{k_2})^2} \frac{w^{+}}{w^{-}-w^{+}} = \frac{1}{\langle \lambda_1 \lambda_2 (w^-) \rangle \langle \bar{\lambda}_1  \bar{\lambda}_2 \rangle} .\label{eq:w-k12l1l2}
\end{align}

Additionally, $\langle \bar{\lambda}_{1} (w^-) \bar{\lambda}_{2}\rangle=0$, implies, $\bar\lambda_{1}(w^-) = C_2 \bar\lambda_{2}$. We know,
\begin{align}
    & \lambda_{1}^{\alpha} \bar{\lambda}_{1}^{\dot{\alpha}} (w) + \lambda_{2}^{\alpha} (w) \bar{\lambda}_{2}^{\dot{\alpha}} =-\lambda_{int}^{\alpha} \bar{\lambda}_{int}^{\dot{\alpha}} , \nonumber \\
 \implies   & (C_2\lambda_{1}^{\alpha} +  \lambda_{2}^{\alpha} (w)) \bar{\lambda}_{2}^{\dot{\alpha}} = -\lambda_{int}^{\alpha} \bar{\lambda}_{int}^{\dot{\alpha}}, \nonumber \\
 \implies & \bar\lambda_{2}= \bar\lambda_{int} \hskip1cm[\text{comparing both sides}] .
\end{align}

For $T(-+++)$, under these conditions, we can see that the nonvanishing contribution to the residue comes from diagram (a) of Fig. \ref{fig:NMHV}.

\subsection{For $p=i(|\textbf{k}_3|+|\textbf{k}_4|)$ pole}
\label{sec:k3k4}

In the case of the pole $p=i(|\bm{k_3}|+|\bm{k_4}|)$, for all three four-point functions considered in this paper,  we have $|\bm{k_3}|+|\bm{k_4}| + ip = 0$. We also know $|\bm{k_3}|+|\bm{k_4}| - ip = E^{34}_{-p}$. This gives, $E^{34}_{-p}=|\bm{k_3}|+|\bm{k_4}| +ip - 2ip=-2ip$. Therefore, we can write,

\begin{align}
 &\lambda_{3}^{'\alpha} \bar{\lambda}_{3}^{'\dot{\alpha}} + \lambda_{4}^{'\alpha}  \bar{\lambda}_{4}^{'\dot{\alpha}} - \lambda_{int}^{\alpha} \bar{\lambda}_{int}^{\dot{\alpha}} =  2 i p\, e^{\alpha \dot{\alpha}} ,\nonumber \\ 
\implies &\lambda_{3}^{'\alpha} \bar{\lambda}_{3}^{'\dot{\alpha}} + \lambda_{4}^{'\alpha}  \bar{\lambda}_{4}^{'\dot{\alpha}} =  \left(\lambda_{int}^{\alpha} \bar{\lambda}_{int}^{\dot{\alpha}} + 2ip e^{\alpha \dot{\alpha}} \right) , \nonumber \\ 
\implies & \left(\lambda_{3}^{'\alpha} \bar{\lambda}_{3}^{'\dot{\alpha}} + \lambda_{4}^{'\alpha}  \bar{\lambda}_{4}^{'\dot{\alpha}}\right) e_{\alpha \beta}e_{\dot{\alpha} \dot{\beta}}\left(\lambda_{3}^{'\beta} \bar{\lambda}_{3}^{'\dot{\beta}} + \lambda_{4}^{'\beta}  \bar{\lambda}_{4}^{'\dot{\beta}}\right) = \left(\lambda_{int}^{\alpha} \bar{\lambda}_{int}^{\dot{\alpha}} + 2ip e^{\alpha \dot{\alpha}} \right) e_{\alpha \beta}e_{\dot{\alpha} \dot{\beta}} \left(\lambda_{int}^{\beta} \bar{\lambda}_{int}^{\dot{\beta}} + 2ip e^{\beta \dot{\beta}} \right) , \nonumber \\
\implies &2 \langle \lambda'_{3} \lambda'_{4} \rangle \langle \bar{\lambda'}_{3} \bar{\lambda'}_{4}\rangle = 0+ 2ip\, e^{\alpha \dot{\alpha}} e_{\alpha \beta}e_{\dot{\alpha} \dot{\beta}} \lambda_{int}^{\beta} \bar{\lambda}_{int}^{\dot{\beta}} + 2ip\,\lambda_{int}^{\alpha} \bar{\lambda}_{int}^{\dot{\alpha}} e_{\alpha \beta}e_{\dot{\alpha} \dot{\beta}} e^{\beta \dot{\beta}} + (2ip)^2\, e^{\alpha \dot{\alpha}} e_{\alpha \beta}e_{\dot{\alpha} \dot{\beta}} e^{\beta \dot{\beta}} , \nonumber \\
\implies &2 \langle \lambda'_{3} \lambda'_{4} \rangle \langle \bar{\lambda'}_{3} \bar{\lambda'}_{4}\rangle = -2\, (2ip)\, \langle \bar{\lambda}_{int} \lambda_{int}\rangle + 2 (2ip)^2 , \nonumber \\
\implies &\langle \lambda'_{3} \lambda'_{4} \rangle \langle \bar{\lambda'}_{3} \bar{\lambda'}_{4}\rangle= 0 \hskip0.5cm[\because \langle \bar{\lambda}_{int} \lambda_{int}\rangle=2ip].
\label{eq:k3k4A}
\end{align}
For the $T(-+++)$ case, particularly, $\lambda'_{3} = \lambda_{3}(w),\,\lambda'_{4} = \lambda_{4}(w),\,\bar\lambda'_{3} = \bar\lambda_{3},\,\bar\lambda'_{4} = \bar\lambda_{4}$. Therefore, from eq. (\ref{eq:k3k4A}) we get, $\langle \lambda_{3}(w) \lambda_{4} (w) \rangle=0$. This implies, $\lambda_{3}(w) = C(w) \lambda_{4}(w)$. Similar to the analysis done for $p=i(|\bm{k_1}|+|\bm{k_2}|)$ pole of $T(++++)$ case, we can express $w^2$ as,
\begin{align}
 w^2 = -\frac{\langle \lambda_{3} \lambda_{4} \rangle}{\beta_{3} \beta_{4} \langle \bar{\lambda}_{3} \bar{\lambda}_{4} \rangle} - \frac{(\beta_3 \langle \bar{\lambda}_{3} \lambda_{4} \rangle  +\beta_4 \langle  \lambda_{3} \bar{\lambda}_{4} \rangle )}{\beta_{3} \beta_{4} \langle \bar{\lambda}_{3} \bar{\lambda}_{4} \rangle} w  . \label{eq:wquadk3k4}
\end{align}

We can also express $C(w)$ in terms of spinor products as, 
\begin{align}
    C(w) = \frac{\langle \bar{\lambda}_3 \lambda_4 \rangle + \beta_4 w \langle \bar{\lambda}_3 \bar{\lambda}_4 \rangle}{2|\bm{k_3}|} \hskip0.5cm[\text{from }\langle \bar{\lambda}_3 \lambda_4(w) \rangle=\langle \bar{\lambda}_3 t(w) \,\lambda_3(w) \rangle] .\label{eq:Cw}
\end{align}

Furthermore, for $p= i (|\bm{k_3}|+|\bm{k_4}|)$, 
\begin{align}
  p^2 + (\bm{k_1} + \bm{k_2})^2 = -\langle \lambda_3 \lambda_4 \rangle \langle \bar{\lambda}_3 \bar{\lambda}_4 \rangle . \label{eq:propk3k4}
\end{align}

The expression for the residue for the $p=i(|\bm{k_3}|+|\bm{k_4}|)$ starting from eq. (\ref{eq:nmhvschan}), after some algebra and using eqs. (\ref{eq:Cw}) and (\ref{eq:propk3k4}), is given by,
\begin{align}
    &\hskip-1cm 2 \pi i \underset{\substack{p=i (|\bm{k_3}|+|\bm{k_4}|)}}{\operatorname{Res}}\,I_s^{(0)} \nonumber \\
    &= \sum_{\pm}  \frac{1}{8  |\bm{k_1}||\bm{k_2}| |\bm{k_3}||\bm{k_4}|} \frac{(|\bm{k_2}| +ip - |\bm{k_1}|)(ip + |\bm{k_1}| -|\bm{k_2}|)}{\langle \lambda_3 \lambda_4 \rangle (|\bm{k_1}|+|\bm{k_2}| + ip)} \nonumber \\
    &\hskip0.5cm \times \left[  \frac{\langle  \lambda_1  \lambda_{int} \rangle^3 \langle  \bar{\lambda}_2  \bar{\lambda}_{int} \rangle^3 \langle \bar{\lambda}_4 \lambda_3 (w)\rangle \langle  \bar{\lambda}_3 \lambda_3 (w)\rangle}{\langle \lambda_1  \lambda_2(w) \rangle \langle  \bar{\lambda}_2  \bar{\lambda}_{int} \rangle^3 \langle \lambda_2 (w) \lambda_{int} \rangle}\right] \hskip-0.1cm \frac{w^{\mp}}{w^{\pm} - w^{\mp}} \Big|_{p=i (|\bm{k_3}|+|\bm{k_4}|)}  \hskip0.2cm[\because \bar{\lambda}_{int}=\lambda_3 (w)]\nonumber \\
     &= \frac{-1}{8  |\bm{k_1}||\bm{k_2}| |\bm{k_3}||\bm{k_4}|} \frac{(|\bm{k_2}|- |\bm{k_3}|-|\bm{k_4}| - |\bm{k_1}|)( |\bm{k_1}| - |\bm{k_3}|-|\bm{k_4}| -|\bm{k_2}|)(|\bm{k_1}|+|\bm{k_2}| -|\bm{k_3}|-|\bm{k_4}|)}{\langle \lambda_3 \lambda_4 \rangle  (|\bm{k_1}| - |\bm{k_2}|+|\bm{k_3}|+|\bm{k_4}| ) } \nonumber \\
    &\hskip0.5cm \times \langle  \lambda_1 \bar{\lambda}_2 \rangle^3 2 |\bm{k_4}| 2 |\bm{k_3}| 2 |\bm{k_3}| \sum_{\pm}  \left[  \frac{ 1}{ (\langle \bar{\lambda}_3 \lambda_4 \rangle + \beta_4 w \langle \bar{\lambda}_3 \bar{\lambda}_4  \rangle)    \langle  \bar{\lambda}_2 \lambda_3 (w) \rangle^2 \langle \lambda_2 (w) \lambda_1 \rangle }\right] \frac{w^{\mp}}{w^{\pm} - w^{\mp}} .  \label{eq:nmhvk3k4A}   
\end{align}

Consider the summed-over part in eq. (\ref{eq:nmhvk3k4A}), i.e., $\sum_{\pm}  \left[  \frac{ 1}{ (\langle \bar{\lambda}_3 \lambda_4 \rangle + \beta_4 w \langle \bar{\lambda}_3 \bar{\lambda}_4  \rangle)    \langle  \bar{\lambda}_2 \lambda_3 (w) \rangle^2 \langle \lambda_2 (w) \lambda_1 \rangle }\right] \frac{w^{\mp}}{w^{\pm} - w^{\mp}}$.  Let the expression in the square brackets be defined as $F(w)$. The denominator of $F(w)$ is a polynomial in $w$ of order 4. We can recursively reduce the order of the polynomial to a linear expression in $w$, using eq. (\ref{eq:wquadk3k4}). $w^4$ can be written as $(w^2)^2$ and $w^3$ can be written as $w(w^2)$, and we repeat the process till we get an expression linear in $w$. We then write $F(w)$ in the form $f_0 + f_1 w$. Using eq. (\ref{eq:fwf0rel}), we know the value of the entire summed-over part is given by $-f_0$, which is,

\begin{align}
    f_0 =  \frac{d_0+d_1W_B}{d_0^{2}+d_0d_1W_B-d_1^{2}W_A} , \label{eq:f0nmhv}
\end{align}
where,
\begin{align}
 W_A&=-\frac{\langle\lambda_3\lambda_4\rangle}
{b_3b_4\langle\bar{\lambda}_3\bar{\lambda}_4\rangle},
\qquad
W_B=-\frac{b_3\langle\bar{\lambda}_3\lambda_4\rangle+b_4\langle\lambda_3\bar{\lambda}_4\rangle}
{b_3b_4\langle\bar{\lambda}_3\bar{\lambda}_4\rangle},\\
d_0&=c_0+c_2W_A+c_3W_AW_B+c_4\big(W_A^{2}+W_AW_B^{2}\big),\\
d_1&=c_1+c_2W_B+c_3\big(W_A+W_B^{2}\big)+c_4\big(2W_AW_B+W_B^{3}\big),\\
c_0&=\frac{1}{2k_3}\,
\langle\bar{\lambda}_3\lambda_4\rangle
\langle\lambda_2\lambda_1\rangle
\langle\bar{\lambda}_2\lambda_3\rangle^{2},
\\[6pt]
c_1&=\frac{1}{2k_3}\Big[
2\beta_3\,
\langle\bar{\lambda}_3\lambda_4\rangle
\langle\lambda_2\lambda_1\rangle
\langle\bar{\lambda}_2\lambda_3\rangle
\langle\bar{\lambda}_2\bar{\lambda}_3\rangle
+\beta_2\,
\langle\bar{\lambda}_3\lambda_4\rangle
\langle\bar{\lambda}_2\lambda_1\rangle
\langle\bar{\lambda}_2\lambda_3\rangle^{2}
\nonumber\\
&\hspace{1.4cm}
+\beta_4\,
\langle\bar{\lambda}_3\bar{\lambda}_4\rangle
\langle\lambda_2\lambda_1\rangle
\langle\bar{\lambda}_2\lambda_3\rangle^{2}\Big],
\\[6pt]
c_2&=\frac{1}{2k_3}\Big[
\beta_3^{2}\,
\langle\bar{\lambda}_3\lambda_4\rangle
\langle\lambda_2\lambda_1\rangle
\langle\bar{\lambda}_2\bar{\lambda}_3\rangle^{2}
+2\beta_2\beta_3\,
\langle\bar{\lambda}_3\lambda_4\rangle
\langle\bar{\lambda}_2\lambda_1\rangle
\langle\bar{\lambda}_2\lambda_3\rangle
\langle\bar{\lambda}_2\bar{\lambda}_3\rangle
\nonumber\\
&\hspace{1.4cm}
+2\beta_3\beta_4\,
\langle\bar{\lambda}_3\bar{\lambda}_4\rangle
\langle\lambda_2\lambda_1\rangle
\langle\bar{\lambda}_2\lambda_3\rangle
\langle\bar{\lambda}_2\bar{\lambda}_3\rangle
+\beta_2\beta_4\,
\langle\bar{\lambda}_3\bar{\lambda}_4\rangle
\langle\bar{\lambda}_2\lambda_1\rangle
\langle\bar{\lambda}_2\lambda_3\rangle^{2}\Big],
\\[6pt]
c_3&=\frac{1}{2k_3}\Big[
\beta_2\beta_3^{2}\,
\langle\bar{\lambda}_3\lambda_4\rangle
\langle\bar{\lambda}_2\lambda_1\rangle
\langle\bar{\lambda}_2\bar{\lambda}_3\rangle^{2}
+\beta_3^{2}\beta_4\,
\langle\bar{\lambda}_3\bar{\lambda}_4\rangle
\langle\lambda_2\lambda_1\rangle
\langle\bar{\lambda}_2\bar{\lambda}_3\rangle^{2}
\nonumber\\
&\hspace{1.4cm}
+2\beta_2\beta_3\beta_4\,
\langle\bar{\lambda}_3\bar{\lambda}_4\rangle
\langle\bar{\lambda}_2\lambda_1\rangle
\langle\bar{\lambda}_2\lambda_3\rangle
\langle\bar{\lambda}_2\bar{\lambda}_3\rangle\Big],
\\[6pt]
c_4&=\frac{\beta_2\beta_3^{2}\beta_4}{2k_3}\,
\langle\bar{\lambda}_3\bar{\lambda}_4\rangle
\langle\bar{\lambda}_2\lambda_1\rangle
\langle\bar{\lambda}_2\bar{\lambda}_3\rangle^{2}.
\end{align}

The $\beta_2$, $\beta_3$, and $\beta_4$ expressions are given in eq. (\ref{eq:betanmhv}).

%For the $\langle-+++\rangle$ case, particularly, similar to  the

For $T(++++)$ and $T(-+-+)$ cases, to obtain the respective residues for the $p=i(|\bm{k_3}|+|\bm{k_4}|)$ pole, we just need to interchange $1\leftrightarrow3$ and $2\leftrightarrow4$ in the respective expressions for residue from $p=i(|\bm{k_1}|+|\bm{k_2}|)$ pole.

%%%%%%%%%%%%%%%%%%%%%%%%%%
\section{Witten Diagram Computation}
\label{sec:witten}
We give the four-point function expressions derived from the Witten diagram computations by \cite{ALM:2021}.

The All-plus case is given by eq. (55) of \cite{ALM:2021}as\footnote{There is a typographical error in eq. (55) of \cite{ALM:2021}. The first spinor product inside the square bracket should be $\langle \lambda_1 \bar{\lambda}_2 \rangle$, instead of $\langle \bar\lambda_1 \bar{\lambda}_2 \rangle$.}, 

\begin{align}
A_{++++} &= \frac{1}{8|\bm{k_1}| |\bm{k_2}| |\bm{k_3}| |\bm{k_4}|}\frac{1}{s} \langle \bar{\lambda}_1 \bar{\lambda}_2 \rangle \langle \bar{\lambda}_3 \bar{\lambda}_4 \rangle \Big[i(\langle \lambda_1 \bar{\lambda}_2 \rangle \langle \bar{\lambda}_4 \bar{\lambda}_1 \rangle \langle \bar{\lambda}_1 \bar{\lambda}_3 \rangle  + \langle \lambda_2 \bar{\lambda}_1 \rangle \langle \bar{\lambda}_3 \bar{\lambda}_2 \rangle \langle \bar{\lambda}_2 \bar{\lambda}_4 \rangle) \nonumber \\
& - |\bm{k_1} + \bm{k_2}| \left(\langle \bar{\lambda}_2 \bar{\lambda}_3 \rangle \langle \bar{\lambda}_4 \bar{\lambda}_1 \rangle - \langle \bar{\lambda}_1 \bar{\lambda}_3 \rangle \langle \bar{\lambda}_2 \bar{\lambda}_4 \rangle \right) - \frac{1}{|\bm{k_1} + \bm{k_2}|} \langle \bar{\lambda}_1 \bar{\lambda}_2 \rangle  \langle \bar{\lambda}_3 \bar{\lambda}_4 \rangle  (|\bm{k_1}| - |\bm{k_2}|)(|\bm{k_3}| - |\bm{k_4}|)\Big] \nonumber \\
&+ 2\leftrightarrow4.
\label{eq:almallp}
\end{align}
 
The Single-minus case is given by eq. (62) of \cite{ALM:2021} as,
\begin{align}
A_{-+++} &= \frac{1}{8|\bm{k_1}| |\bm{k_2}| |\bm{k_3}| |\bm{k_4}|}\frac{1}{s}\langle \lambda_1 \bar{\lambda}_2 \rangle \langle \bar{\lambda}_3 \bar{\lambda}_4 \rangle\Big[i (\langle \lambda_1 \lambda_2 \rangle \langle \bar{\lambda}_2 \bar{\lambda}_4 \rangle \langle \bar{\lambda}_2 \bar{\lambda}_3 \rangle + \langle \bar{\lambda}_2 \bar{\lambda}_1 \rangle \langle \lambda_1 \bar{\lambda}_3 \rangle \langle \lambda_1 \bar{\lambda}_4 \rangle) \nonumber \\
&+(|\bm{k_1} + \bm{k_2}|+2|\bm{k_1}|) (\langle \lambda_1 \bar{\lambda}_4 \rangle \langle \bar{\lambda}_2 \bar{\lambda}_3 \rangle - \langle \lambda_1 \bar{\lambda}_3\rangle \langle \bar{\lambda}_4 \bar{\lambda}_2 \rangle)  \nonumber \\
&-\frac{1}{|\bm{k_1} + \bm{k_2}|} \langle \lambda_1 \bar{\lambda}_2 \rangle \langle \bar{\lambda}_3 \bar{\lambda}_4 \rangle (|\bm{k_1}| - |\bm{k_2}|)(|\bm{k_3}| - |\bm{k_4}|) \nonumber \\
&+\frac{2i |\bm{k_1}| \langle \lambda_1 \bar{\lambda}_3 \rangle \langle \lambda_1 \bar{\lambda}_4 \rangle}{t}\left(\langle \bar{\lambda}_1 \bar{\lambda}_2 \rangle (E + 2|\bm{k_2} + \bm{k_3}|) + i \langle \lambda_4 \bar{\lambda}_2 \rangle \langle \bar{\lambda}_4 \bar{\lambda}_1 \rangle \right)\Big] + 2 \leftrightarrow 4. \label{eq:almnmhv}
\end{align}

The MHV case is given by eq. (66) of \cite{ALM:2021} as,
\begin{align}
A_{-+-+} &= \frac{1}{8|\bm{k_1}| |\bm{k_2}| |\bm{k_3}| |\bm{k_4}|}\frac{1}{E\,s}\langle \lambda_1 \bar{\lambda}_2 \rangle \langle \lambda_3 \bar{\lambda}_4 \rangle \Big[4( |\bm{k_1}| |\bm{k_4}| + |\bm{k_2}| |\bm{k_3}|)\langle \lambda_1 \lambda_3 \rangle \langle \bar{\lambda}_2 \bar{\lambda}_4 \rangle \nonumber \\
&+ i (E - 2|\bm{k_2}| - 2|\bm{k_4}|)(\langle \lambda_1 \lambda_2 \rangle \langle \bar{\lambda}_2 \bar{\lambda}_4 \rangle \langle \lambda_3 \bar{\lambda}_2 \rangle + \langle \bar{\lambda}_2 \bar{\lambda}_1 \rangle \langle \lambda_3 \lambda_1 \rangle \langle \lambda_1 \bar{\lambda}_4 \rangle) \nonumber \\
&+ E |\bm{k_1} + \bm{k_2}| (\langle \lambda_1 \lambda_3 \rangle \langle \bar{\lambda}_2 \bar{\lambda}_4 \rangle - \langle \lambda_1 \bar{\lambda}_4 \rangle \langle \lambda_3 \bar{\lambda}_2 \rangle) \nonumber \\
&-\frac{1}{|\bm{k_1} + \bm{k_2}|}E \langle \lambda_1 \bar{\lambda}_2 \rangle \langle \lambda_3 \bar{\lambda}_4 \rangle (|\bm{k_1}| - |\bm{k_2}|)(|\bm{k_3}| - |\bm{k_4}|)\Big] + 2\leftrightarrow4.
\label{eq:almmhv}
\end{align}

In the above equations $s=(|\bm{k_1}|+|\bm{k_2}|+|\bm{k_1}+\bm{k_2}|)(|\bm{k_3}|+|\bm{k_4}|+|\bm{k_3}+\bm{k_4}|)$ and
$t=(|\bm{k_1}|+|\bm{k_4}|+|\bm{k_1}+\bm{k_4}|)(|\bm{k_3}|+|\bm{k_2}|+|\bm{k_3}+\bm{k_2}|)$.

%%%%%%%%%%%%%%%%%%%%%%%%%%
\end{appendix}
%%%%%%%%%%%%%%%%%%%%%%%%%%

\bibliographystyle{JHEP}
\bibliography{references}

\end{document}